\documentclass[journal]{IEEEtran}

\usepackage[colorlinks]{hyperref}
\usepackage{graphics} 
\usepackage{amsmath}
\usepackage{amssymb}  
\usepackage{amsfonts} 
\usepackage{graphicx,float,wrapfig}
\usepackage{cite}
\usepackage{bm}
\usepackage{soul,color}
\usepackage[linesnumbered,ruled]{algorithm2e}
\usepackage{url}
\usepackage{booktabs}
\usepackage[table,xcdraw]{xcolor}
\usepackage{multirow}
\usepackage[]{changes}
\usepackage{MnSymbol}
\usepackage{marginnote}

\newcommand{\argmin}{\text{argmin}}

\newcommand{\bmat}{\begin{bmatrix}}
\newcommand{\emat}{\end{bmatrix}}
\newcommand{\rf}{\text{ref}}

\usepackage{lipsum}

\definecolor{ao(english)}{rgb}{0.0, 0.5, 0.0}
\newtheorem{proposition}{Proposition}

\begin{document}
%
\title{Lane-Change in Dense Traffic with \\Model Predictive Control and Neural Networks}
%
%
%

\author{Sangjae~Bae,~\IEEEmembership{Member,~IEEE,}
        David~Isele,~\IEEEmembership{Member,~IEEE,}
        Alireza~Nakhaei,
        Peng~Xu,~\IEEEmembership{Member,~IEEE,}
        Alexandre~Miranda~A$\tilde{\text{n}}$on,
        Chiho~Choi,
        Kikuo~Fujimura,~\IEEEmembership{Member,~IEEE,}
        Scott~Moura,~\IEEEmembership{Member,~IEEE,}
\thanks{S. Bae was with University of California, Berkeley, CA, 95134 USA, at the time of the work, and is currently with Honda Research Institute, CA, 95134 USA e-mail: {\tt sbae@honda-ri.com}.}
\thanks{D. Isele, A. Miranda Anon, C. Choi, K. Fujimura are with Honda Research Institute, CA, 95134 USA e-mail: {\tt\{disele, amiranda, cchoi, kfujimura\}@honda-ri.com}.}
\thanks{A. Nakhaei was with Honda Research Institute, CA, 95134 USA, at the time of the work, and is currently with Toyota Research Institute, CA, 94022 USA e-mail: {\tt alireza.nakhaei@tri.global}.}%
\thanks{P. Xu was with Honda Research Institute, CA, 95134 USA, at the time of the work, and is currently with Amazon, WA, 98109 USA e-mail: {\tt xpharry@amazon.com}.}%
\thanks{S. Moura is with University of California, Berkeley,
CA, 94720 USA e-mail: {\tt smoura@berkeley.edu}.}
}

%
%

\markboth{IEEE TRANSACTIONS ON CONTROL SYSTEMS TECHNOLOGY}%
{Shell \MakeLowercase{\textit{et al.}}: Bare Demo of IEEEtran.cls for IEEE Journals}
%



\maketitle

\begin{abstract}
This paper presents an online smooth-path lane-change control framework. We focus on dense traffic where inter-vehicle space gaps are narrow, and cooperation with surrounding drivers is essential to achieve the lane-change maneuver. We propose a two-stage control framework that harmonizes Model Predictive Control (MPC) with Generative Adversarial Networks (GAN) by utilizing driving intentions to generate smooth lane-change maneuvers. To improve performance in practice, the system is augmented with an adaptive safety boundary and a Kalman Filter to mitigate sensor noise. Simulation studies are investigated in different levels of traffic density and cooperativeness of other drivers. The simulation results support the effectiveness, driving comfort, and safety of the proposed method.
\end{abstract}


%
\IEEEpeerreviewmaketitle

\section{Introduction}\label{sec:introduction}
%
%
%
%
\IEEEPARstart{G}{iven} the millions of edge cases related to navigating traffic situations safely and efficiently, 
designing a reliable motion planner is still an open-ended challenge. In this study, we focus on one of the extreme, yet often observed traffic conditions: highly dense traffic. 

Advanced Driver Assistance Systems (ADAS), such as Adaptive Cruise Control (ACC) and braking assistance, have effectively enhanced safety \cite{neale2005overview,peden2004world} by mitigating the impact of human errors and also improving energy efficiency \cite{guanetti2018control, bae2022ecological}. A large body of literature in the domain of ADAS has focused on lane-change assistance that involves three stages of the decision making process \cite{nilsson2016if}: (i) \textit{WHERE} to change lanes (if desired), (ii) \textit{WHEN} to change lanes, and (iii) \textit{HOW} to change lanes. Decision-making at each stage then engages unique technical challenges and an ample set of literature does exist in ordinary driving conditions. For \textit{WHERE}, if a lane-changing is discretionary, a target lane is often chosen by comparing an average speed or throughput of neighboring lanes with that of a current lane \cite{nilsson2016if, sun2011lane, rahman2013review}. If a neighboring lane has a higher throughput, lane-change is desired and the neighboring lane is set as a target lane. For \textit{WHEN}, inter-vehicle gaps are often investigated to find a safe (target) gap to execute lane-change \cite{schildbach2015scenario}. For \textit{HOW}, given a target lane and target gap, a control trajectory is found to execute a smooth lane-change \cite{chen2013lane, kelly2003reactive}. In highly dense traffic, however, the stages are not clearly classified. That is, \textit{HOW} and \textit{WHEN} are closely combined since there is no safe inter-vehicle gap and the gap can be created by how the vehicle is controlled, i.e., \textit{WHEN} is determined by \textit{HOW}. Therefore, to succeed at changing lanes in dense traffic, it is important to estimate other vehicles' behaviors conditioned on how the ego vehicle\footnote{We refer to the vehicle that is controlled as the ``ego vehicle''. } behaves. Corresponding technical challenges of lane-changing in dense traffic include:
\renewcommand{\labelenumi}{\textbf{{[C\arabic{enumi}]}}}
\begin{enumerate}
    \item the absence of safe inter-vehicle gap;
    \item the need of evaluating ``cooperative'' behaviors triggered by a complex decision mechanism;
    \item the presence of multiple vehicles simultaneously playing a role;
    \item the practicality of \textit{online} computation.
\end{enumerate}

\begin{figure}
    \centering
    \includegraphics[width=0.9\columnwidth]{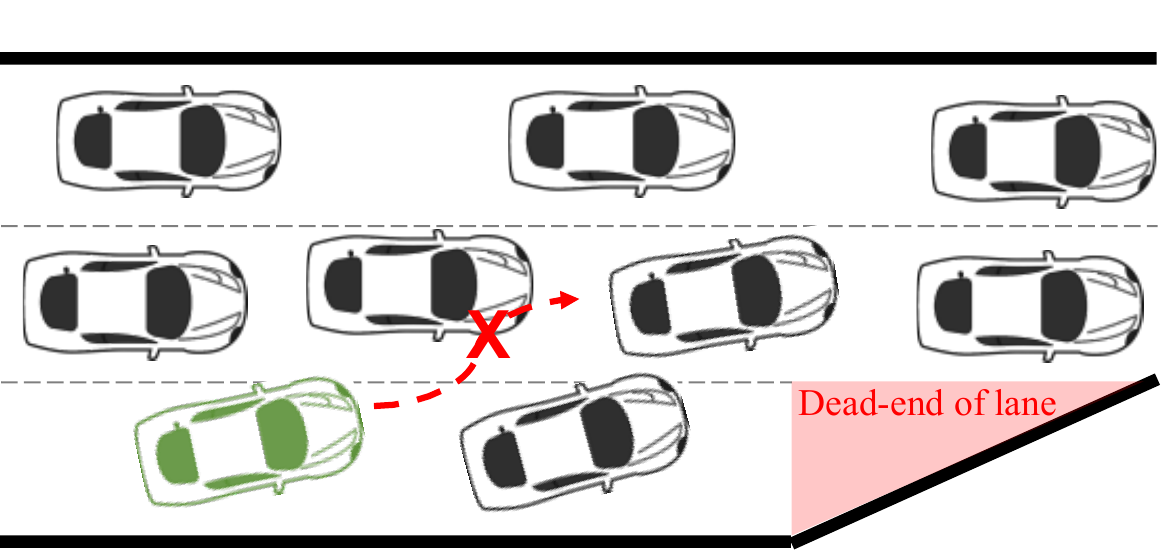}
    \caption{The autonomous-driving vehicle (in \textcolor{ao(english)}{green}) intends to change lanes, within a restricted merging area. The traffic is dense with narrow inter-vehicle intervals that are spatially insufficient for a vehicle to merge into. The autonomous-driving vehicle would get stuck in the merging area, unless other drivers slow down to make space for the vehicle. }
    \label{fig:motivation}
\end{figure}
An increasing number of literature on lane-changing in dense traffic addresses the above challenges either partially \cite{bouton2019cooperation, naranjo2008lane, you2015trajectory, sadigh2016planning, hu2019IDAS, isele2019interactive} or fully \cite{bae2019cooperation, saxena2019driving, lee2022spatiotemporal}. One recognized study \cite{sadigh2016planning} utilizes Inverse Reinforcement Learning (RL) to estimate a reward for a neighboring vehicle being cooperative. The estimated reward is then used to optimize controls. Recently, a model-free RL agent is designed \cite{saxena2019driving} for lane-changing control in dense traffic, proving its effectiveness in learning complex interactive-behavior mechanisms of multiple surrounding vehicles. Still, reliability and interpretability limit the practical use of RL particularly in safety-critical applications like autonomous driving, and these issues remain active research areas \cite{garcia2015comprehensive, alshiekh2018,isele2018safe}.

Apart from RL, an increasing amount of literature has applied Deep Neural Network architectures to autonomous driving for explaining complex environments \cite{tian2018deeptest, chen2015deepdriving, bojarski2017explaining}. Particularly in predicting motions of humans (drivers), Recurrent Neural Network (RNN) architectures have rapidly advanced, proving their accuracy \cite{choi2019drogon, alahi2016social} as well as computational efficiency \cite{gupta2018social}. Therefore, it would be natural to take advantage of those advances for controller design, in combination with control-theoretic methods for autonomous driving control \cite{doyle2013feedback} and well-formulated vehicle dynamics models \cite{Kong2015}. That is, a controller based on control theory determines optimal control inputs for autonomous driving, while partially exploiting predictions by RNN. The incorporation thereby would keep the controller interpretable, and tunable, while having the controller responsive to delicate, interactive motions of human drivers. It is still challenging to mathematically incorporate RNN into formal controller design and to solve the control problem effectively and efficiently. We address these challenges in this paper.

We add two original contributions to the literature: (i) We propose a mathematical control framework that systematically evaluates several human drivers' interactive motions, in dense traffic. The framework is built upon Model Predictive Control (MPC) and integrates Recurrent Neural Networks (RNN) as a \textit{sub-model} -- which we denote as ``NNMPC'' for convenience. RNN exclusively predicts future positions of other vehicles, reflecting interactions between vehicles. Then the MPC framework evaluates the predictions through a formalized optimization structure. (ii) We propose a \textit{real-time} numerical algorithm that finds optimal solutions based on driving-intentions. The idea of incorporating RNN as a prediction model into an MPC controller is straightforward. However, a challenge with RNNs is their highly nonlinear structure, which is complex to solve using classical optimization algorithms. We show our algorithm generates smooth lane-change maneuvers while securing a \textit{real-time} computational efficiency. Additionally, we enhance the practicality by accounting for sensor noises, prediction errors, and recursive feasibility. We also adapt a commonly-used two-stage structure of planning \& control and validate the proposed framework under the CARLA simulation environment.

This paper incorporates and extends our previous work in  \cite{bae2019cooperation}. In particular, we develop an action space refinement technique utilizing driving intentions, which improves driving comfort and computational efficiency. Besides, this paper develops the theoretical model of our previous work into a robust system more capable of meeting the demands of a physical system. Finally, the proposed model is compared with other cooperation-aware lane-changing models, including \cite{isele2019interactive}, to validate its performance.

The paper is organized in the following manner: Section~\ref{sec:framework} gives an overview of the two-stage planning and control framework, followed by detailed formulations and algorithms of the proposed trajectory planner in Section~\ref{sec:planner}. Section~\ref{sec:simulation} reports the performance of the proposed method in the CARLA simulation environment. The paper concludes with a summary in Section~\ref{sec:conclusion}. 

\section{Overview of Structure and Model}\label{sec:framework}

\subsection{Structure}
\begin{figure}
    \centering
    \includegraphics[width=.8\columnwidth]{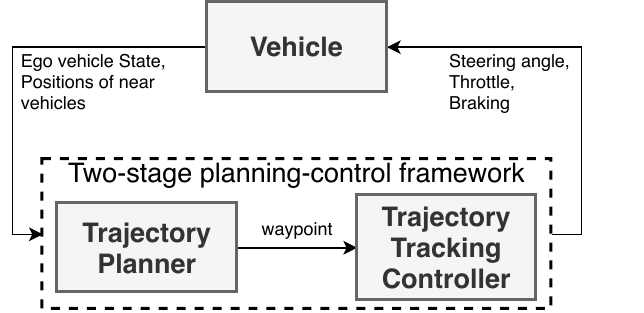}
    \caption{Diagram for two-stage planning and control framework. The trajectory planner in the upper layer generates waypoints, a sequence of positions coupled with velocities. The trajectory tracking controller in the lower layer determines a pair of steering angle and throttle/braking. The trajectory planner evaluates cooperative behaviors conditioned on possible choice of trajectory.}
    \label{fig:diag-planning-control}
\end{figure}

In our previous work \cite{saxena2019driving}, we proposed a control framework that combines trajectory planning with determining control inputs, i.e., one-stage. This one-stage structure, however, may not be appropriate where a low-level controller already exists and true system plants are barely known. Additionally, the one-stage structure limits the flexibility in tuning the planning and control as they are coupled. Therefore, in many practical scenarios, a two-stage structure is utilized \cite{polack2017kinematic,febbo2020accurate}, where the upper layer plans trajectories (planner), and the lower layer follows the trajectories (controller). Fig.~\ref{fig:diag-planning-control} illustrates the two-stage structure, which we leverage in this work. The planner in the upper layer (bottom left box in Fig.~\ref{fig:diag-planning-control}) builds an MPC formulations that incorporate RNN as a prediction module. The controller in the lower layer (bottom right box in Fig.~\ref{fig:diag-planning-control}) consists of two separate control process for longitudinal control and lateral control, respectively. This separation enables ease in implementations. The longitudinal control is canonically designed as a proportional-integral-derivative (PID) controller, and the lateral control is designed as an MPC controller \cite{tashiro2013vehicle}. This paper focuses on the upper layer, i.e., trajectory planner while briefly summarizing the lower layer controller design for completeness of the work.

\subsection{Controller}
The longitudinal controller is a PID controller that takes a desired speed and current speed to compute an offset and devises a throttle in the next time step according to PID gains tuned \textit{a priori}. 

The lateral controller takes a reference position (from waypoints) in next time step and current speed as input. It finds a steering angle that minimizes a divergence between ``next'' and ``propagated'' position through the vehicle rotation model in \cite{tashiro2013vehicle}. The MPC formulation reads:
\begin{align}
\delta(t) = \argmin_{\delta}\; &(x^\rf (t+1)-\tilde{x}(t+1))^2 \nonumber\\&+ (y^\rf (t+1)-\tilde{y}(t+1))^2 \nonumber\\&+ \lambda_\psi (\psi^\rf (t+1)-\tilde{\psi}(t+1))^2,
\end{align}
subject to the vehicle rotation model, where $\tilde{}$ denotes a predicted value and superscript ${}^\rf$ denotes a reference value. The parameter $\lambda_\psi$ relatively weights the penalty on heading. Recall, designing a lower-level controller is not the main focus of this paper. Hence, the design is motivated by simplicity. More advanced lower-level controllers can be formulated, e.g., combining longitudinal control with lateral control \cite{attia2014combined}. 

\subsection{Vehicle Model}\label{ssec:system_dynamics}
For trajectory planning, we utilize the nonlinear kinematic bicycle model in \cite{Kong2015kinematic} to represent the vehicle dynamics. For completeness, we re-write the kinematics here:
\begin{align}
    \dot{x}&=v\cos(\psi+\beta)\\
    \dot{y}&=v\sin(\psi+\beta)\\
    \dot{\psi}&=\frac{v}{l_r}\sin(\beta)\\
    \dot{v}&=a\\
    \beta&=\tan^{-1}\left(\frac{l_r}{l_f+l_r}\tan(\delta)\right)
\end{align}
where ($x$,$y$) is the Cartesian coordinate for the center of vehicle, $\psi$ is the inertial heading, $v$ is the vehicle speed, $a$ is the acceleration of the car's center in the same direction as the velocity, and $l_f$ and $l_r$ indicate the distance from the center of the car to the front axles and and to the rear axles, respectively. The control inputs are: (front wheel) steering angle $\delta$ and acceleration $a$. We use Euler discretization to obtain a discrete-time dynamical model in the form:
\begin{equation}
    z(t+1) = f(z(t),u(t)),
\end{equation}
where $z = \begin{bmatrix}x&y&\psi&v\end{bmatrix}^\top$ and $u = \begin{bmatrix}a&\delta\end{bmatrix}^\top$ for time $t$. Note that the above simple bicycle kinematics can improve computational efficiency without losing modeling accuracy \cite{Kong2015kinematic}, compared to more complex models, e.g., vehicle dynamics with tire models \cite{carvalho2013predictive, gao2010predictive}.

\section{Trajectory Planning}\label{sec:planner}
\begin{figure}
    \centering
    \includegraphics[width=1\columnwidth]{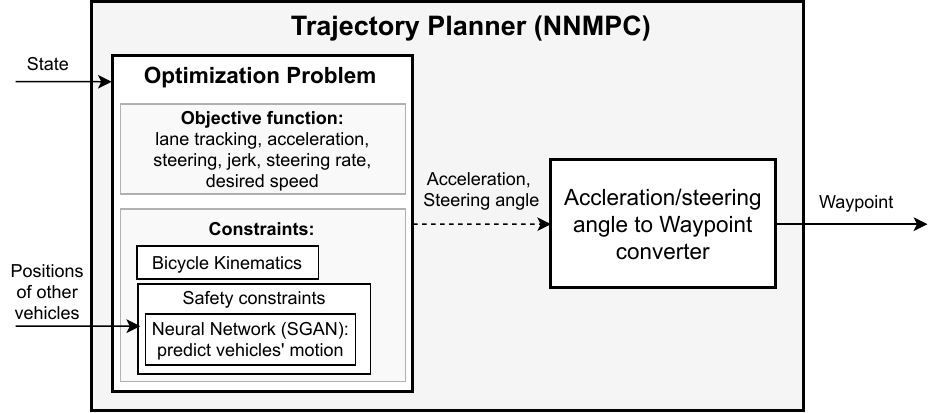}
    \vspace{-10pt}
    \caption{Diagram of the planning framework with a Recurrent Neural network, SGAN. The optimization is solved with respect to accelerations and steering angles which are then converted to waypoints that the lower level controller in Fig.~\ref{fig:diag-planning-control} tracks. }
    \label{fig:diag-nnmpc}
    \vspace{-10pt}
\end{figure}

The trajectory planner is composed of two problems: (i) smooth trajectories generations and (ii) optimality and safety evaluations. The first problem is to find a set of smooth trajectories that reflect different driving intentions (safety is not considered). Then the second problem is to evaluate both the optimality and safety of the pre-computed trajectories (neural networks are applied conditioned on the trajectories). In other words, the first problem (referred to as the pre-computation problem) pre-computes a set of trajectories and the second problem (referred to as the optimization problem) chooses the best one among the trajectories. This hierarchical structure is due to the complexity of the optimization problem that limits both finding solutions itself and computing in real-time. We detail each problem in the following sections.

\subsection{Pre-computation of Smooth Trajectories Reflecting Driving Intentions}\label{sec:intention_trajectory}
\begin{figure}
    \centering
    \includegraphics[trim={0mm 2mm 0mm 2mm}, clip,width=0.8\columnwidth]{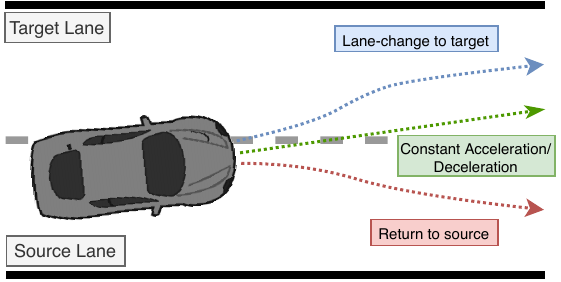}
    \caption{Examples of driving intentions. 
    }
    \label{fig:driving_intention}
    \vspace{-15pt}
\end{figure}

Driving in urban areas takes advantage of the presence of lanes. Lanes are used as guidelines to limit the number of intentions available to the vehicle. We specify four intentions: (i) keeping on a current (source) lane, (ii) changing to a next (target) lane, (iii) constantly increasing speed, (iv) and constantly decreasing speed. For each intention, a trajectory is pre-computed and converted to a sequence of accelerations/braking coupled with steering angles. Examples of pre-computed trajectories are illustrated in Fig.~\ref{fig:driving_intention}. 
It is important to highlight that each trajectory needs to be computed timely for online control while remaining feasible with respect to control limits. 
To compute trajectories for (i) keeping on a source lane or (ii) moving to a target lane, we utilize polynomial spiral curves \cite{kelly2003reactive}. Polynomial spiral curves are commonly used in path-planning problems for their curvature continuity which facilitates smooth-path generation and for their computational efficiency. We briefly review the path-planning problem with cubic spiral curves \cite{kelly2003reactive}, where a curve $\mathcal{K}(l)$ is formulated: 
\begin{subequations}
\begin{equation}
    \mathcal{K}(l) = \beta_3 l^3 + \beta_2 l^2 + \beta_1 l + \beta_0,
\end{equation}
for arbitrary distance step $l$. Positions $x,y$ and heading $\psi$ are formulated:
\vspace{-10pt}
\begin{align}
    x(l)&=x_0 + \int^{l}_0 \cos(\psi(l^\prime)) dl^\prime,\\
    y(l)&=y_0 + \int^{l}_0 \sin(\psi(l^\prime)) dl^\prime,\\
    \psi(l)&=\psi_0 + \int^{l}_0 \mathcal{K}(l^\prime) dl^\prime.
\end{align}
\end{subequations}
Given boundary conditions for positions and heading, numerical approximations using Simpson's rule (for integrals) \cite{atkinson2008introduction}, and parameter mappings (between spiral parameters $\beta$ and curvature $p$), the smooth-path optimization problem reads:
\begin{subequations}
\begin{align}
\min_{p}\; &f_{be}(\beta_0,\beta_1,\beta_2,\beta_3,l_f) \nonumber\\&+ \lambda_x(x_s([p]_4)-x_f) + \lambda_y (y_s([p]_4)-y_f) \nonumber\\&+ \lambda_\psi(\psi_s([p]_4) - \psi_f),
\end{align}

subject to:
\begin{align}
    |[p]_1| &\leq \mathcal{K}_{\text{max}},\label{eq:curv_p1}\\
    |[p]_2| &\leq \mathcal{K}_{\text{max}},\label{eq:curv_p2}
\end{align}\label{eq:smooth_path}
\end{subequations}
where $p$ indicates a vector of curvatures with $[p]_i$ indicating the $i^{th}$ element of $p$. Subscript $s$ denotes an approximation with Simpson's rule, and subscript $f$ denotes a final point. The first term in the objective function is the penalty on bending energy, formulated as:
\begin{equation}
f_{be}(\beta_0,\beta_1,\beta_2,\beta_3,l_f) = \int_0^{l_f}\;\mathcal{K}(l^\prime)^2 dl^\prime,
\end{equation}
and the second through the last term in the objective function respectively represent softened boundary conditions (i.e., soft constraints) for final positions $(x_f,y_f)$ and heading $\psi_f$. The penalty weights $\lambda_x, \lambda_y$, and $\lambda_\psi$ on the soft constraints are arbitrarily set to a high value. Note that the curvature constraints \eqref{eq:curv_p1},\eqref{eq:curv_p2} are applied only to a couple points, which effectively reduces a number of variables to optimize\footnote{Limiting the number of curvature constraints, the cubic spiral curve still generates curvatures the vehicle can perform \cite{kelly2003reactive}.}. The maximum curvature $\mathcal{K}_{\text{max}}$ is set to the maximum steering angle $\delta_{\text{max}}$, i.e., $\mathcal{K}_{\text{max}}=\delta_{\text{max}}$. The optimization problem is non-convex and non-linear with respect to the curvature $p$. To solve the problem, we apply L-BFGS-B \cite{zhu1997algorithm}. Recall that a resulting path is continuous and only a few equally-spaced points\footnote{The number of points is equivalent to the number of time steps in the receding horizon $T$} are sampled among the path to evaluate its optimality and feasibility of lane-changing in Section \ref{sec:planning_obj}. It is important to note that the pre-computed trajectories are converted to control sequences of acceleration $a$ and steering angle $\delta$.

\subsection{Objective}\label{sec:planning_obj}
The planning objective is to merge to the target lane while avoiding collisions with other vehicles. We prefer to change lanes sooner than later. We also prefer smooth accelerations and steering for passenger comfort. The objective function is formulated:
\begin{align}
    J&=\sum_{\ell=t}^{t+T}\lambda_{div}(\ell|t)D(\ell|t)
    &&\textbf{target lane}
    \label{eq:obj_divergence}\\
     &+\sum_{\ell=t}^{t+T}\lambda_{v}\|v(\ell|t)-v^{\text{ref}}\|^2
     &&\textbf{target velocity}\\
     &+\sum_{\ell=t}^{t+T-1}\lambda_{\delta}\|\delta(\ell|t)\|^2 
     &&\textbf{steering effort}
     \label{eq:obj_front_wheel_angle}\\
     &+\sum_{\ell=t}^{t+T-1}\lambda_{a}\|a(\ell|t)\|^2
     &&\textbf{accel. effort}\label{eq:obj_acc}\\
     &+\sum_{\ell=t+1}^{t+T-1}\lambda_{\Delta\delta}\|\delta(\ell|t)-\delta(\ell-1|t)\|^2
     &&\textbf{steering rate}\label{eq:obj_front_wheel_angle_jerk}\\
     &+\sum_{\ell=t+1}^{t+T-1}\lambda_{\Delta a}\|a(\ell|t)-a(\ell-1|t)\|^2
     &&\textbf{jerk}\label{eq:obj_jerk}
\end{align}
where $(\ell|t)$ indicates time $\ell$ based on the measurements at time $t$. Symbol $x_\text{end}$ is the latitude coordinate of the road-end, $D(\ell|t)$ is the distance norm for the vector between the ego vehicle's center and the target lane at time $\ell$, $v^{\text{ref}}$ is the reference velocity. Each penalty is regularized with $\lambda_{div}$, $\lambda_v$, $\lambda_{\delta}$, $\lambda_{a}$, $\lambda_{\Delta \delta}$, and $\lambda_{\Delta a}$, respectively. We incentivize a timely lane change with the dynamic weight $\lambda_{div}(\ell|t) = \frac{1}{\sqrt{(x(\ell|t)-x_\text{end})^2+(y(\ell|t)-y_\text{end})^2}}$. As the ego vehicle gets closer to the deadend position $(x_\text{end},y_\text{end})$, the dynamic weight exponentially increases. 
The term \eqref{eq:obj_divergence} penalizes the divergence of the center of the vehicle from the vertical center of the target lane. The term \eqref{eq:obj_front_wheel_angle} and \eqref{eq:obj_acc} penalize the control effort of steering angle and acceleration, respectively. The term \eqref{eq:obj_front_wheel_angle_jerk} and \eqref{eq:obj_jerk} penalize the steering rate and jerk, respectively, for drive comfort. 

\subsection{Safety Constraints with a Recurrent Neural Network}
To quantify safety, we consider a distance measure between two vehicles. That is, if the distance between two vehicles is zero, it means they collide with each other. Now, there are two important aspects to set up a safety constraint. First, a mathematical measure of distance between two vehicles depends on how each vehicle is shaped since we do not have direct measurements of inter-vehicle distances over a planning horizon -- recall, we have a measurement only at the current time, not every time step over a planning horizon. Second, the distance must be calculated based on predicted positions with respect to the ego vehicle's trajectories. Namely, we need a prediction module that estimates positions of other vehicles in response to the ego vehicle's actions -- and we leverage a recurrent neural network.

\subsubsection{Minimum Inter-vehicle Distance Measure}
\begin{figure}
    \centering
    \includegraphics[width=0.6\columnwidth]{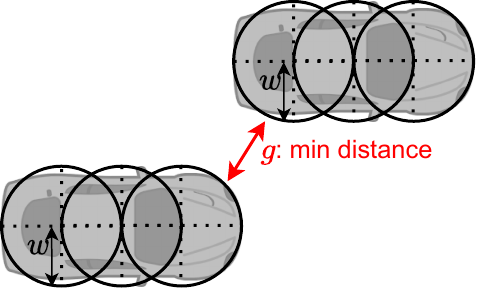}
    \caption{Vehicle shapes are modeled by three circles. Note that multiple (more than three) circles of various radius can be applied to any vehicle shape and size. The minimum distance $g$ is computed as the minimum Euclidean distance between any pair of circle, as in Eqn. \eqref{eq:const_circle}}
    \label{fig:vehicle_shape}
    \vspace{-10pt}
\end{figure}

As illustrated in Fig.~\ref{fig:vehicle_shape}, we first model the vehicle shape with three circles where the radius is half the vehicle width. Then an Euclidean distance between two circles are analytically obtained (which is computationally efficient). To constrain safety, a minimum distance between any pairs of circles must be greater than a safety bound $\epsilon$. Formally:
\begin{equation}
     g_i(x,y)=\min_{p,q\in\{-1,0,1\}} d_{i}(p,q) \geq \epsilon,\label{eq:const_circle}
\end{equation}
\begin{align}
    &\text{where: } d_{i}(p,q)\nonumber\\ 
    &= \Big[\big((x+p(h-w)\cos\psi)-(x_i+q(h_i-w_i)\cos\psi_i)\big)^2\nonumber\\
    &\;\;+\big((y+p(h-w)\sin\psi)-(y_i+q(h_i-w_i)\sin\psi_i)\big)^2\nonumber\Big]^{\frac{1}{2}}\\
    &\;\;-(w+w_i)\label{eq:dist_circle},
\end{align}
and $w$ and $h$ are, respectively, half width and half height of the vehicle, respectively, and $\epsilon$ is a safety bound. Subscript $i$ indicates vehicle $i$. Note that Eqn. \eqref{eq:dist_circle} represents a distance between two circles. 

\subsubsection{Interactive Motion Prediction}
Drivers' motions are responsive to interactions with each other, and consequently, the motions must be predicted simultaneously. We adopt a state-of-the-art RNN structure, Social Generative Adversarial Networks (SGAN) from \cite{gupta2018social} which efficiently and effectively captures interactions between agents (drivers). SGAN is composed of a generator and discriminator that are adversarial to each other. Both the generator and discriminator are comprised of long-short term memory (LSTM) networks to account for the sequential nature of agents' motion. The following module aggregates the agents' motion states to evaluate their interactions. This pooling process is essential to share the motion history between agents, generating social interactions as a pooled tensor $P_i$ for \textit{each} agent $i$. The decoder in the generator then predicts multiple trajectories that are socially interactive with each other. SGAN takes as an input a sequence of positions for each agent within a scene over an observation time horizon $T_{obs}$. It outputs a sequence of positions for each agent over a prediction time horizon $T_{pred}$. Interested readers are referred to \cite{gupta2018social} for more details. It is important to highlight that a trained SGAN will predict the most probable reactions of other vehicles based on the control commands on the ego vehicle, and the history of previous actions. However, in reality, the reactions of other vehicles might be different.

\subsubsection{Incorporation of Recurrent Neural Network}
A trained SGAN predicts the evolution of vehicle $i$'s centroid. These predictions are incorporated into safety constraints \eqref{eq:const_circle}. Formally, we denote a trained SGAN as a function $\phi(t)$ that maps observed trajectories to predicted trajectories for $N$ nearby vehicles at time $t$:
\begin{align}
    &\phi(t)\;:\;\nonumber\\ 
    &\begin{bmatrix}
    (x_1(t),y_1(t))&\cdots&(x_{N}(t),y_{N}(t))\\
    \vdots&\vdots&\vdots\\
    \begin{matrix}(x_1(t-T_{obs}+1),\\\qquad y_1(t-T_{obs}+1))\end{matrix} &\cdots&\begin{matrix}(x_{N}(t-T_{obs}+1),\\\qquad y_{N}(t-T_{obs}+1))\end{matrix}
    \end{bmatrix}\nonumber\\
    &\vspace{3mm}\hspace{4cm}{\downmapsto}\vspace{3mm}\nonumber\\
    &\quad\begin{bmatrix}
    (\hat{x}_1(t+1),\hat{y}_1(t+1))&\cdots&(\hat{x}_{N}(t+1),\hat{y}_{N}(t+1))\\
    \vdots&\cdots&\vdots\\
    \begin{matrix}(\hat{x}_1(t+T_{pred}),\\\qquad \hat{y}_1(t+T_{pred}))\end{matrix}&\cdots&\begin{matrix}(\hat{x}_{N}(t+T_{pred}),\\\qquad\hat{y}_{N}(t+T_{pred}))\end{matrix}
    \end{bmatrix},
\end{align}
where $\hat{\cdot}$ indicates a predicted value. Given the observations until time $t$, the coordinates of vehicle $i$ at time $t+1$ are represented as $\hat{x}_{i}(t+1) = \phi_{i,x}(t)$ and $\hat{y}_{i}(t+1) = \phi_{i,y}(t)$. Note that the trained SGAN is not time dependent, i.e., the notation $(t)$ in $\phi(t)$ indicates the evaluation at time $t$ without any indication of temporal dependencies.

\subsection{Optimization Formulations}
The complete optimization problem for the receding horizon control is:
\begin{subequations}
\begin{align}
    \min_{z,a,\delta}\;\; J&=\sum_{\ell=t}^{t+T}\Big(\lambda_{div}(x(\ell|t);x_{\text{end}})D(\ell|t)\nonumber\\
     &\quad\quad\quad+\lambda_v\|v(\ell|t)-v^{\text{ref}}\|^2\Big)\nonumber\\
     &+\sum_{\ell=t}^{t+T-1}\Big(\lambda_{\delta}\|\delta(\ell|t)\|^2+\lambda_{a}\|a(\ell|t)\|^2\Big)\nonumber\\
     &+\sum_{\ell=t+1}^{t+T-1}\Big(\lambda_{\Delta\delta}\|\delta(\ell|t)-\delta(\ell-1|t)\|^2\nonumber\\
     &\quad\quad\quad+\lambda_{\Delta a}\|a(\ell|t)-a(\ell-1|t)\|^2\Big)
\end{align}\label{eq:obj}
subject to:
\begin{align}
    z(\ell+1|t)&=f(z(\ell|t),\delta(\ell|t),a(\ell|t))\label{eq:const_dynamics}\\
    g_i (z(\ell+1|t);x_i(t),y_i(t)) &\geq \epsilon, \quad\forall i \in \{1,\cdots,N_{veh}\}\label{eq:const_collision}\\
    a(\ell|t) &\in [a_{\text{min}},a_{\text{max}}]\label{eq:const_a}\\
    \delta(\ell|t) &\in [\delta_{\text{min}},\delta_{\text{max}}]\label{eq:const_delta}\\
    (x(t+T|t),y(t+T|t)) &\in \mathcal{S} \label{eq:const_x_end}
\end{align}\label{eq:opt}
\end{subequations}
where $(\ell|t)$ denotes time $\ell$ based on the measurements at time $t$, $z$ is a stacked variable of states, $a$ denotes acceleration, $\delta$ denotes steering angle, and $D(\ell|t)$ is the distance norm for the vector between the ego vehicle's center and the target lane at time $\ell$. The objective \eqref{eq:obj} assesses reference tracking, control effort, and drive comfort. Each penalty in \eqref{eq:obj} is weighted by $\lambda$ as described in Section~\ref{sec:planning_obj}. The constraint \eqref{eq:const_dynamics} indicates the nonlinear bicycle kinematics \cite{Kong2015kinematic}. The constraint \eqref{eq:const_collision} indicates collision avoidance, enforcing the Euclidean distance between the ego vehicle and vehicle $i$ to be larger than the safety bound $\epsilon$. Controls $(a,\delta)$ are lower and upper bounded in \eqref{eq:const_a} and \eqref{eq:const_delta}. The constraint \eqref{eq:const_x_end} ensures the ego vehicle not to get stuck at a road-end positioned at $(x_\text{end},y_\text{end})$ where $\mathcal{S}$ denotes a feasible set that avoids freezing behaviors. We detail the constraint for avoiding freezing behaviors in the next section. 

\subsubsection{Terminal Constraint for Avoiding Freezing Problem}\label{subsec:recursive_feas}
\indent To prevent the ego vehicle being stuck at the dead end, we define the feasible set of a terminal position as a half-plane in the two-dimensional Euclidean space:
\begin{align}
    \mathcal{S}_\theta = \{(x^\prime, y^\prime)\;|\;&y^\prime \cos \theta - x^\prime \sin \theta \nonumber\\&\geq f_{\text{aff}}(y^\prime\sin\theta+x^\prime\sin\theta)\}, \label{eq:feas_set}
\end{align}
where:
\begin{align}
    f_{\text{aff}}(x) = \delta_\text{max}(x-x_\text{end})+y_\text{end}, \label{eq:affine_function}
\end{align}
with road angle $\theta$. Note that bicycle kinematics represents an ellipsoidal trajectory (with a constant velocity and steering angle) and the set bounded by the ellipsoidal trajectory is a convex set. Therefore, by definition, the affine function \eqref{eq:affine_function} is enveloped by the trajectory between any two points on the trajectory. Each subset split by the affine function still represents a convex set and the volume of any subset is less than that of the primal set \cite{boyd2004convex}. In other words, regardless of the road angle and lane-changing direction, the feasible set \eqref{eq:feas_set} with the affine function \eqref{eq:affine_function} represents a more conservative set of trajectories than that with the bicycle kinematics between two points on the trajectory.

\subsubsection{Adaptive Safety with Prediction Errors}\label{sec:sgan_safety}
Prediction errors of neural networks are often evaluated with Average Displacement Error (ADE) computed as a Euclidean distance between ``predicted'' and ``actual'' positions over a prediction horizon. The ADE can be used as a distance buffer in the safety constraint \eqref{eq:const_collision} to mitigate the prediction error. The ADE from training, however, may differ from testing and therefore the ADE needs to be adaptively evaluated in real-time. Therefore, in real-time, we evaluate a maximum prediction error among all subject vehicles' position at each time and add additional distance buffer in longitudinal and lateral directions, respectively. Namely:
\begin{align}
    \epsilon^\text{lng}_i&=\left(\max_{i=1,\cdots,N_{\text{veh}}}\text{ADE}_i\right)\cos{\psi_i},\nonumber\\
    \epsilon^\text{lat}_i&=\left(\max_{i=1,\cdots,N_{\text{veh}}}\text{ADE}_i\right)\sin{\psi_i},\label{eq:safety_bound}
\end{align}
where:
\begin{equation}
    \text{ADE}_i = \sum_{\ell=t-T+1}^{t}\frac{\|(\hat{x}_i(\ell),\hat{y}_i(\ell))-(x_i(\ell),y_i(\ell))\|_2}{T},
\end{equation}
$(\hat{x}_i(t),\hat{y}_i(t))$ is a predicted position of vehicle $i$ at time step $t$. The real-time prediction error is then applied in longitudinal and lateral directions, respectively. That is, the safety constraint \eqref{eq:const_collision} now reads:

\begin{align}
    \min_{p,q\in\{-1,0,1\}} d_{i}(p,q)\cos\psi_i &\geq \epsilon_0+\epsilon^\text{lng}_i,\\
    \min_{p,q\in\{-1,0,1\}} d_{i}(p,q)\sin\psi_i &\geq \epsilon_0+\epsilon^\text{lat}_i,
\end{align}
where $\epsilon_0$ is a default distance buffer. Note that the real-time prediction error is applied in longitudinal and lateral directions, respectively, to avoid over-conservative behaviors (e.g., pushing brake when a vehicle is on the next lane as the lateral distance is within the safety margin).
Note that the distance buffer can apply distributionally robust optimization methods \cite{kandel2020distributionally} to provably improve safety, which remains for future work.

\subsection{Complete Algorithm}
\begin{figure}
    \centering
    \includegraphics[width=0.9\columnwidth]{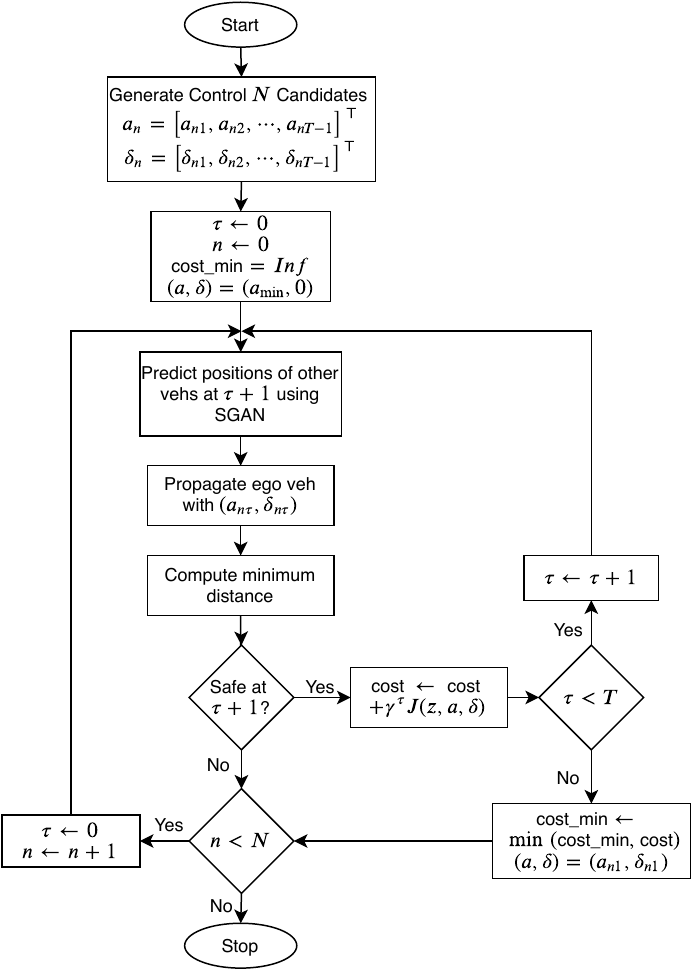}
    \caption{Finding optimal controls of NNMPC. At each time step $\tau$, (i) predict other drivers' position (for the next time step, i.e., $\tau+1$), (ii) propagate the ego vehicle with a pre-computed control at time $\tau$ through the dynamics \eqref{eq:const_dynamics}, (iii) check the constraints \eqref{eq:const_collision}-\eqref{eq:const_x_end}, and discards the candidate sequence if violated, and otherwise (iv) update the cumulative cost \eqref{eq:obj} with discounting factor $\gamma \in [0,1]$. Repeat (i)-(iv) for all control candidates and return the candidate that has a minimum cost. If no candidate is feasible, the ego vehicle takes the maximum brake with zero steering.Note, the choice of action when no feasible candidate exists can be extended depending on traffic and road, e.g,. forced returning back to the source lane at high-speed.}
    \label{fig:nnmpc_alg_diag}
\end{figure}

The complete Intention-based NNMPC algorithm with localization and perception filtering is structured in Algorithm~\ref{alg:i_nnmpc}. The algorithm proceeds in the following manner. Suppose a target lane is given. The state of the ego vehicle and the positions of other drivers are initialized with initial measurements. While the divergence $D$ of the ego vehicle's position from the target lane is larger than a minimum distance threshold $\epsilon_\text{min}$, control sequences are pre-computed by \eqref{eq:smooth_path}. Then each control sequence is evaluated by \eqref{eq:opt} (detailed in Fig.~\ref{fig:nnmpc_alg_diag}). With an optimal control sequence, waypoints are generated and sent to a lower-level controller. Meanwhile, new states of the ego vehicle and positions of other drivers are measured and updated.

\begin{algorithm}
    \begin{small}
    \SetKwInOut{Input}{Input}
    \SetKwInOut{Output}{Output}
    \SetKwInOut{Init}{Init}
    \Init{states $z = z_0$, \\
    other vehicles' position $(x_i,y_i) = (x_{i0},y_{i0})$ for all $i\in\{1,\cdots,N_{veh}\}$}
    \While{$D \leq \epsilon_{\text{min}}$}{
        Generate control candidates for different driving intentions
        $\mathcal{U}=\{U_{\text{target}}, U_{\text{source}}, U_{\text{acc}}, U_\text{dcc}\}$\label{alg:gen_controls}\\
        \vspace{3mm}
        
        Find the optimal sequence that minimizes cumulative cost over $T$ and that is feasible with \eqref{eq:const_collision} (Fig.~\ref{fig:nnmpc_alg_diag})\\
        $U_\ast \leftarrow \argmin_{U\in\mathcal{U}}\; \eqref{eq:obj}$ \label{alg:eval_cost}\\
        \vspace{3mm}
        
        Propagate through dynamics \eqref{eq:const_dynamics} with $U_\ast$ to generate waypoints\\
        $z^{\text{ref}} \leftarrow f(z,U_\ast)$\label{alg:monte_update_state}\\
        \vspace{3mm}
        
        Observe state of ego vehicle at the current time step $t$\\ 
        $z \leftarrow \hat{z}$ 
        \vspace{3mm}
        
        Observe positions of other vehicles at the current time $t$\\ 
        $(x_i,y_i) \leftarrow (\hat{x}_i(t),\hat{y}_i(t))$ for all $i$
    }
    \caption{Intention-based NNMPC algorithm}\label{alg:i_nnmpc}
    \end{small}
\end{algorithm}\vspace{-15pt}

\vspace{10pt}
\section{Simulations}\label{sec:simulation}
\subsection{Simulation Setup}
\begin{figure}
    \centering
    \includegraphics[width=0.8\columnwidth]{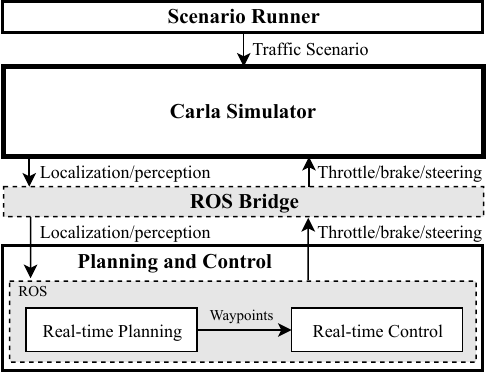}
    \caption{Simulation Setup. The upper three components (Scenario runner, CARLA simulator, and ROS bridge) are open source. 
    }
    \label{fig:sim_setup}
\end{figure}

Shown in Fig.~\ref{fig:sim_setup}, the simulation setup is composed of three main components: (i) CARLA simulator \cite{dosovitskiy2017carla}, (ii) Scenario runner, and (iii) Planning and control. The CARLA simulator is an open-source simulation platform for autonomous driving research, which has gained significant attention due to its scalability with vehicles/pedestrians, flexibility in traffic scenarios, and practicality in implementations with Robot Operating Systems (ROS). The Scenario runner is also an open-source script that enables customization of traffic scenarios in CARLA simulations. The planning and control components represent the proposed methods. We run the simulations on Ubuntu 16.04 LTS (Intel Xeon CPU ES-2640 v4 @2.40GHz x 20, GeForce GTX TITAN).

\subsection{Training Networks}\label{sec:training_sgan}
\subsubsection{Driver Model} \label{ssec:driver_model}
\begin{figure}
    \centering
    \includegraphics[width=.8\columnwidth,trim={0 1.2cm 0 1.1cm},clip]{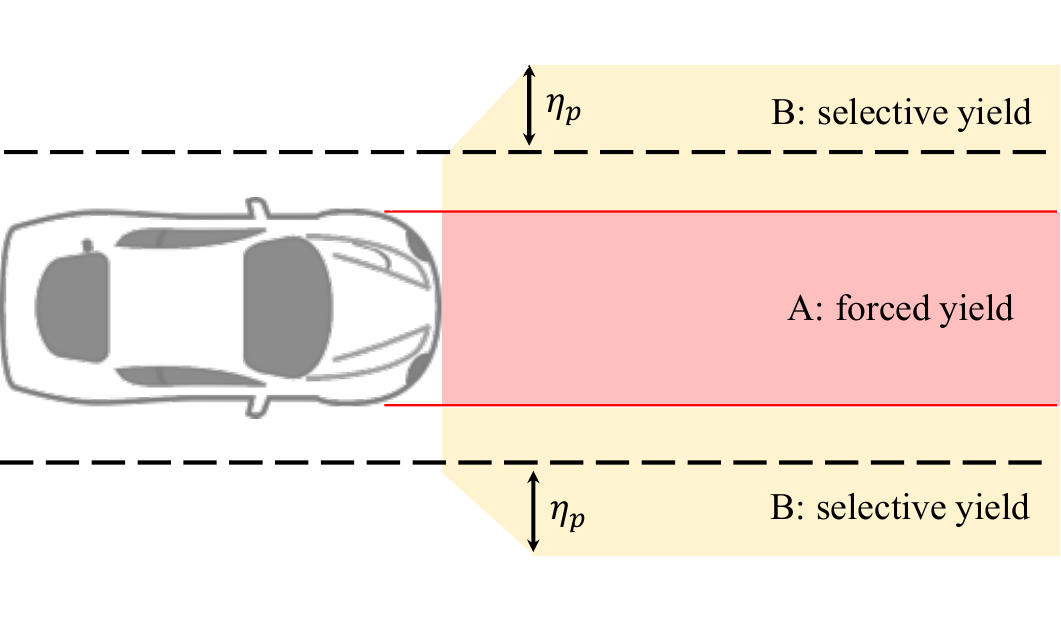}
    \caption{\small{Forced yield zone (red area labeled with A) and selective yield zone (yellow area labeled with B). The dashed lines indicate the boundaries of the center lane. If a vehicle from the next lane intersects with the path of the vehicle, i.e. in zone A, the vehicle in the center lane must stop and wait until the other vehicle cuts into the center lane. If the other vehicle intersects with zone B, then the vehicle in the center lane decides to either yield or not, according to the cooperativeness parameter $\eta_{c}$. Zone B corresponds to a vertical perception range, and the range can be adjusted by $\eta_{\textit{p}}$.}}
    \label{fig:driver_model_yield_zone}
    \vspace{-10pt}
\end{figure}

We train SGAN through simulations. We first define a driver model in which the longitudinal dynamics are controlled by an intelligent driver model (IDM) from \cite{treiber2000congested}. The lane-changing behavior is governed by the strategy of Minimizing Overall Braking Induced by Lane changes (MOBIL) from \cite{kesting2007general}. The driver model is also based on the bicycle kinematics in Section \ref{ssec:system_dynamics}. Additionally, we introduce a parameter for cooperativeness $\eta_{c}\in [0,1]$ to the driver model. If $\eta_{c} = 1$, a vehicle stops and waits until another vehicle within the selective yield zone (zone B in Fig.~\ref{fig:driver_model_yield_zone}) overtakes. If $\eta_{c} = 0$, then the vehicle ignores neighboring vehicles and drives forward. If $0 < \eta_{c} < 1$, then the yield action is randomly sampled from the Bernoulli distribution with probability $p=\eta_{c}$.

\begin{figure}
    \centering
    \includegraphics[width=1\columnwidth,trim={0cm 0.1cm 0.5cm 1.1cm},clip]{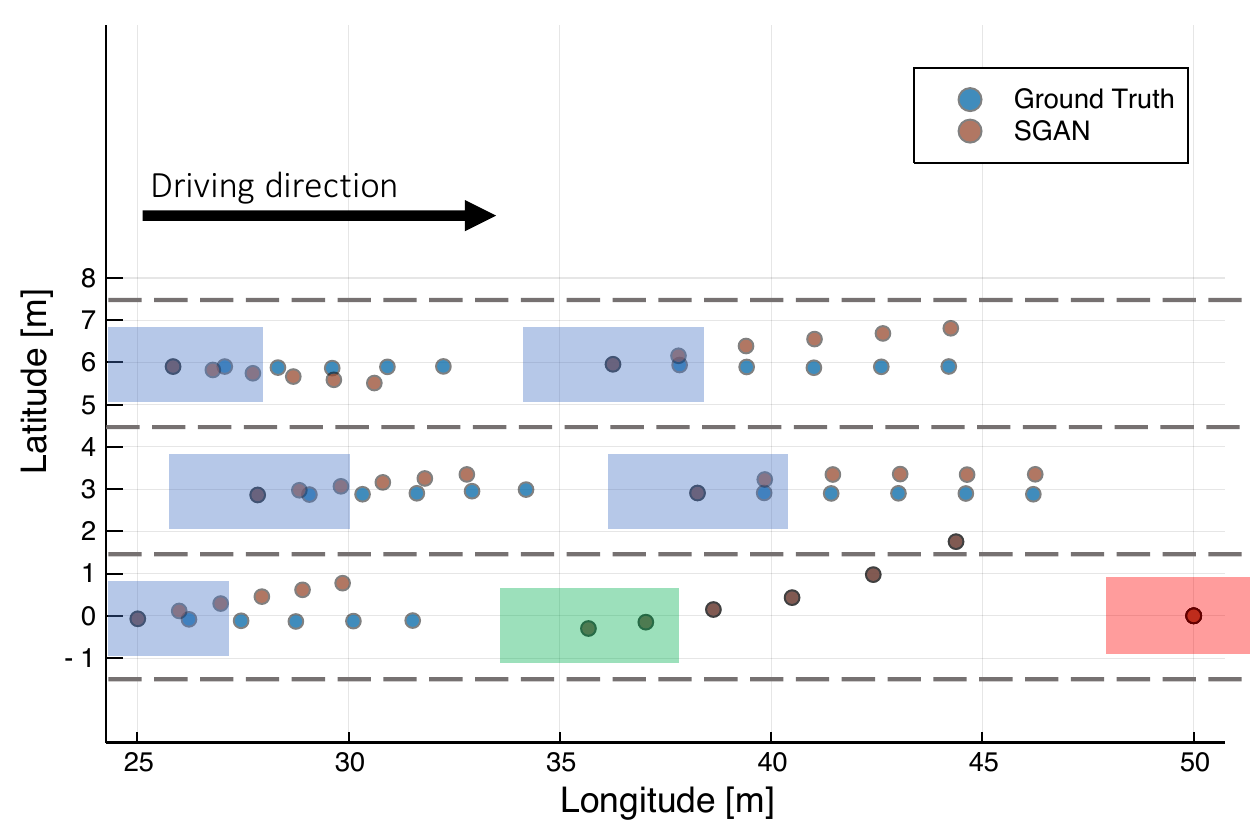}
    \caption{Motions predicted by SGAN. Rectangles in green, blue, and red indicate the ego vehicle, other vehicles, and the stopped vehicles respectively. Dashed lines represent the lane boundary. Each circular point indicates the vehicle center at each time step.}
    \label{fig:sgan_prediction}
    \vspace{-10pt}
\end{figure}

TABLE~\ref{table:sgan_params} lists hyperparameters for the SGAN trained through simulations. Training and validation datasets are generated by simulations with only the driver model from Section \ref{ssec:driver_model}, with heterogeneous parameters from TABLE~\ref{table:driver_params}. The dataset is collected from multiple scenarios in various traffic densities, from free flow to dense traffic. In the dataset, we also add noise to the positions so that the trained neural network becomes more robust to noisy perceptions. With a total of 27550 data points, training the SGAN with a GPU takes approximately 18 hours. An example of motions predicted by SGAN, compared to ground truth, is illustrated in Fig.~\ref{fig:sgan_prediction}. 

\begin{table}[]
\caption{{SGAN parameters}}
\vspace{-5pt}
\begin{tabular}{@{}lll@{}}
\toprule
\multicolumn{1}{c}{Param} & \multicolumn{1}{c}{Description} & \multicolumn{1}{c}{Value} \\ \midrule
$T_{obs}$ & Observation time horizon & $8$ \\
$T_{pred}$ & Prediction time horizon & $2$ \\
$N_{b}$ & Batch size & $64$ \\
$D_{emb}$ & Embedding dimension & $64$ \\
$D_{mlp}$ & MLP dimension & $256$ \\
$D_{G,e}$ & Hidden layer dimension of encoder (generator) & $32$ \\
$D_{G,d}$ & Hidden layer dimension of decoder (generator) &  $64$ \\
$D_{D,e}$ & Hidden layer dimension of encoder (discriminator) & $64$ \\
$D_{bot}$ & Bottleneck dimension in Pooling module & $1024$ \\
$\alpha_G$ & Generator learning rate & $5\cdot10^{-4}$ \\
$\alpha_D$ & Discriminator learning rate & $5\cdot10^{-4}$ \\ \bottomrule
\end{tabular}
\label{table:sgan_params}
\end{table}

\begin{table}[]
\caption{{Driver model design parameters}}
\begin{tabular}{@{}lll@{}}
\toprule
\multicolumn{1}{c}{Param} & \multicolumn{1}{c}{Description} & \multicolumn{1}{c}{Value} \\ \midrule
$\tilde{v}^{\text{ref}}$ & Reference velocity $[m/s]$ & $\mathcal{U}(2, 5)$ \\
$\tilde{T}$ & Safe time headway $[s]$ & $\mathcal{U}(1, 2)$ \\
$\tilde{a}_{\text{max}}$ & Maximum acceleration $[m/s^2]$ & $\mathcal{U}(2.5, 3.5)$ \\
$\tilde{b}$ & Comfortable deceleration $[m/s^2]$ & $\mathcal{U}(1.5, 2.5)$ \\
$\tilde{\delta}$ & Acceleration exponent & $\mathcal{U}(3.5, 4.5)$ \\
$\tilde{s}_0$ & Minimum distance to front vehicle $[m]$ & $\mathcal{U}(1, 3)$ \\
$\eta_{c}$ & Cooperativeness $\in [0,1]$ & $\mathcal{U}(0, 1)$ \\
$\eta_{p}$ & Vertical perception range $[m]$ & $\mathcal{U}(-0.15, 0.15)$ \\
$w$ & Length from center to side of vehicle $[m]$ & 0.9 \\
$h$ & Length from center to front of vehicle $[m]$ & 2 \\ \bottomrule
\end{tabular}
\label{table:driver_params}
\end{table}

\begin{figure}
    \centering
    \includegraphics[width=1\columnwidth]{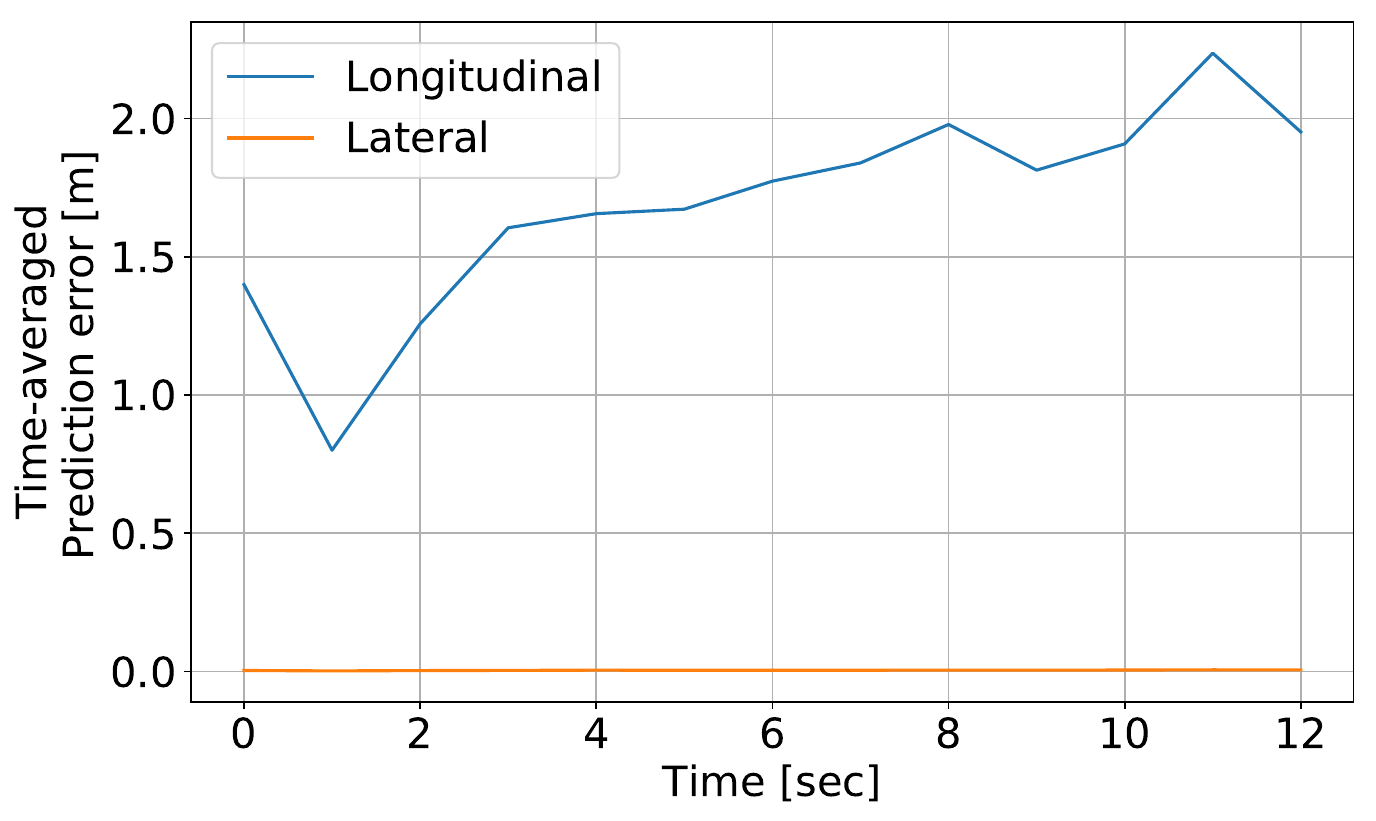}
    \vspace{-10pt}
    \caption{Real-time (time-averaged) prediction errors in longitudinal and lateral directions. The prediction in lateral direction is very accurate while that in longitudinal direction is not as accurate. To mitigate the prediction error, the safety margin $\epsilon$ in \eqref{eq:const_collision} is set accordingly in each longitudinal and lateral direction. Note that the mean is 1.703 $[m]$, which is comparable with the training error of the SGAN (1.872 $[m]$).}
    \label{fig:realtime_prediction_error}
\end{figure}

The SGAN has 1.872 $[m]$ of average displacement error and 2.643 $[m]$ of final displacement error after $T_{pred}=2$, for the training data. It is important to note that the training dataset does not include the testing scenarios. Figure~\ref{fig:realtime_prediction_error} shows the \textit{real-time} time-averaged errors that are comparable with the aforementioned training error in the specific testing scenario (discussed in the next section). The advantage of the adaptive safety boundary (based on the real-time prediction errors) is not obvious against the static safety boundary, however, it should not be underestimated especially under unknown conditions where the real-time observation can enhance safety. Recall, the prediction of SGAN is obviously not perfect. Still, we leverage its capability of evaluating interactive behaviors, which is the key to changing lanes under highly dense traffic. The prediction errors are embedded into the safety bound \eqref{eq:safety_bound}.
Also, our ultimate goal is a safe lane change, as discussed next, not ``zero-error'' motion prediction.

Note that SGAN can provide a distribution of predicted motions, which can be incorporated into the optimization problem as chance constraints, thereby enabling a robust formulation \cite{kandel2020distributionally}. Also, different methods for designing loss functions for SGAN training can be applied, which are topics for future work.

\subsection{Testing Scenario}
\begin{figure}
    \centering
    \includegraphics[width=0.8\columnwidth]{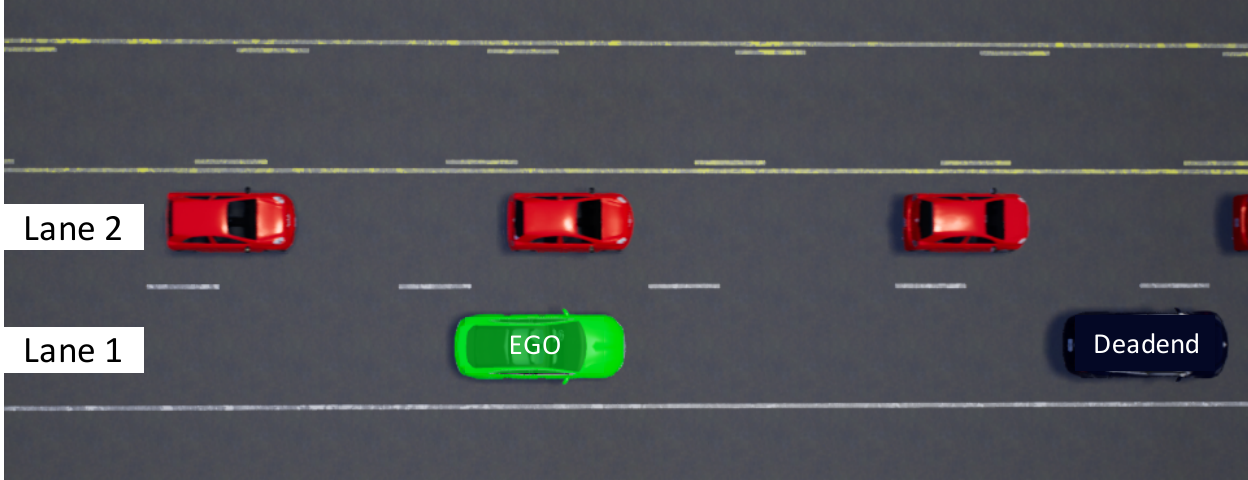}
    \caption{Lane-changing scenario in CARLA. The \textbf{\textcolor{green}{green}} vehicle is the ego vehicle and the \textbf{\textcolor{black}{black}} vehicle is a broken car. The \textbf{\textcolor{red}{red}} vehicles are other traffic participants. Lane 1 is the source lane and Lane 2 is the target lane. The objective of the ego vehicle is to lane-change to the target lane before the dead end.}
    \label{fig:carla-top-down}
    \vspace{-15pt}
\end{figure}

Figure~\ref{fig:carla-top-down} illustrates a traffic scenario. Only the ego vehicle applies the proposed method. The other drivers are modeled as IDM and their parameters are randomly sampled from uniform distributions as in TABLE~\ref{table:driver_params}. We examine four traffic scenarios with two different levels of density and cooperativeness: (i) sparse-cooperative, (ii) sparse-aggressive, (iii) dense-cooperative, and (iv) dense-aggressive. The average initial inter-vehicle gap is $10.0\;[m]$ with $1.75\;[s]$ of time headway in sparse scenarios and $7.75\;[m]$ with $0.875\;[s]$ of time headway in dense scenarios. Note that both sparse and dense scenarios do not have a ``free space'' where the ego vehicle can merge-in without interacting with other vehicles (as we target scenarios that require cooperation).
The cooperativeness is measured as the likelihood of slowing down when another vehicle is perceived to change lanes. In aggressive scenarios, other drivers do not slow down unless another vehicle overlaps the trajectory. Zero mean Gaussian noises are added to localization and perception to simulate sensor noise. For localization, by default, the standard deviation (std) in (longitudinal position, lateral position) is set to $(0.3,0.1)\;[m]$ and in heading is $0.08\;[rad]$. For perceptions, by default, std is $(0.1,0.1)\;[m]$ in positions and $0.08\;[rad]$ in heading.

\subsection{Simulation Results}
\begin{figure*}
\centering
\includegraphics[width=.4\columnwidth]{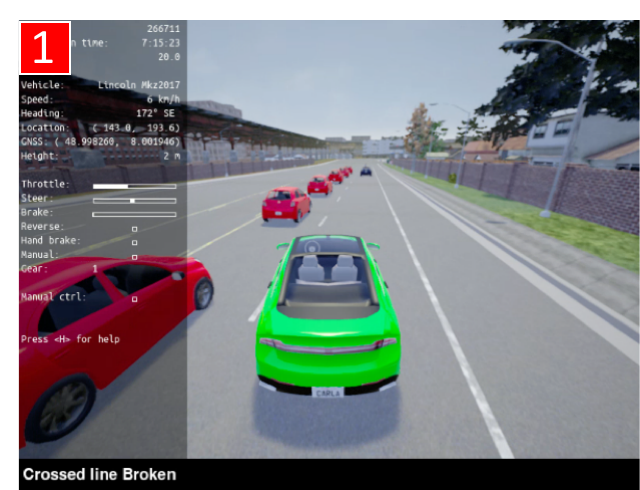}
\includegraphics[width=.4\columnwidth]{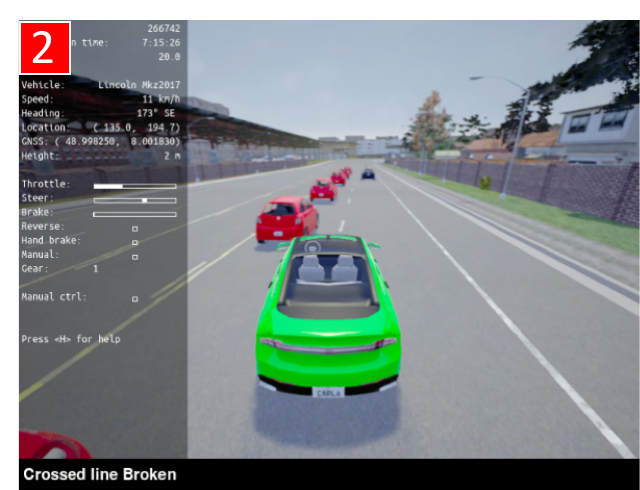}
\includegraphics[width=.4\columnwidth]{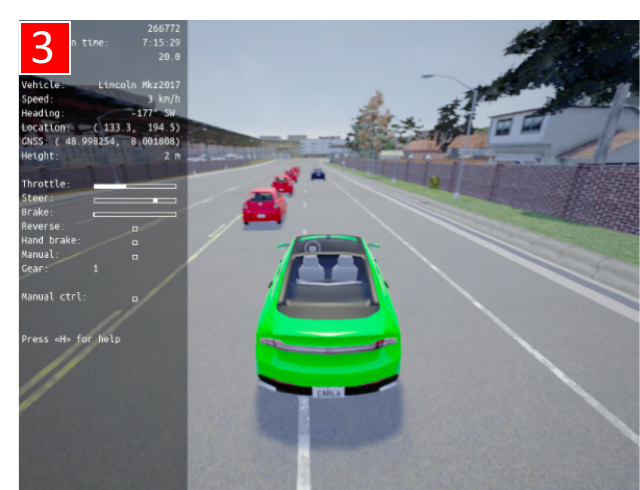}
\includegraphics[width=.4\columnwidth]{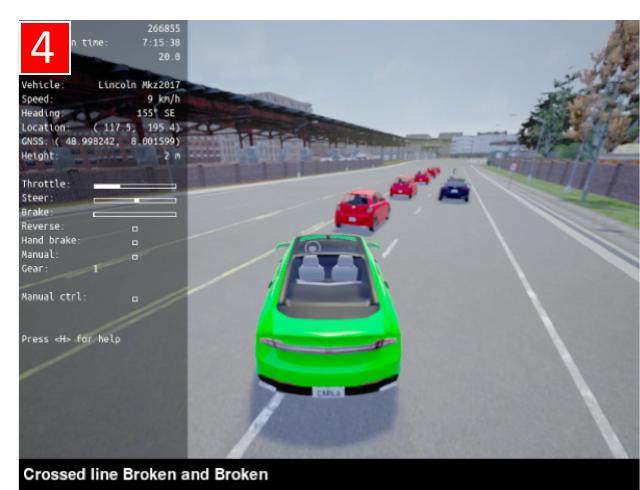}
~
\includegraphics[width=.4\columnwidth]{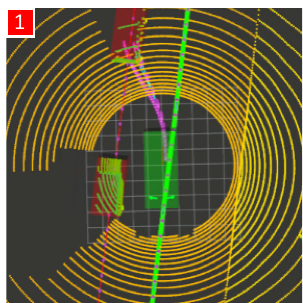}
\includegraphics[width=.4\columnwidth]{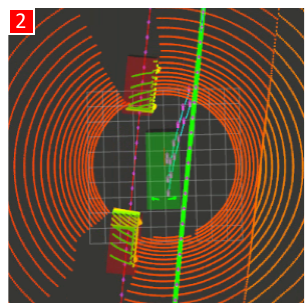}
\includegraphics[width=.4\columnwidth]{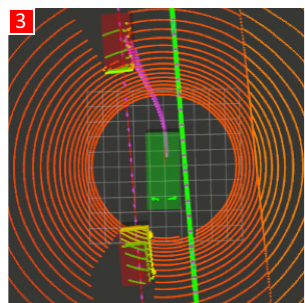}
\includegraphics[width=.4\columnwidth]{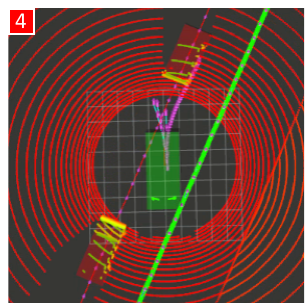}
\caption{CARLA simulation snapshots. The top row illustrates third-person view of the ego vehicle (\textcolor{green}{green}). The bottom row shows lidar scans, source lane (green line), target lane (red line), and waypoints (purple dots connected by cyan lines). Each snapshot indicates, in order, that the ego vehicle: (1) tries to lane-change to target lane, expecting cooperative behavior of the rear vehicle, (2) returns to source lane as the rear vehicle gets too close without slowing down, (3) again tries to lane-change to target lane as the rear vehicle starts slow down, and (4) successfully changes lane.}\label{fig:carla_result}
\vspace{-10pt}
\end{figure*} 
\subsubsection{Qualitative Analysis}
In most cases, NNMPC successfully finds lane-change maneuver and one common procedure is observed: (i) the ego vehicle gets close to the target lane; (ii) a rear vehicle on the target lane slows down; (iii) the ego vehicle changes the lane. The success in lane-change is mainly led by accurate predictions of whether other drivers are being cooperative or not. However, predictions can often be inaccurate, and consequently, the maneuver fails to induce other drivers to slow down. In these instances, the NNMPC maneuvers the ego vehicle back to the source lane and tries to lane-change again. See the demonstrations in Fig.~\ref{fig:carla_result} (video available at {\tt https://github.com/bsj1216/NNMPC}). To measure the optimality of the trajectory, we compare an average jerk and steering rate. Simulation studies show that utilizing driving intentions to generate a smooth-path yields a small average jerk $(0.14\;[\frac{m}{s^2}])$ and average steering rate $(0.37\;[\frac{rad}{s^2}])$.

\subsubsection{Quantitative and Comparative Analysis}
For the four traffic scenarios with different levels of density and cooperativeness, we examine a quantitative analysis to validate the proposed method. For comparative analysis, we consider two baseline methods: (i) a game-theoretic framework that estimates driving intentions of other drivers using a \textit{game tree} \cite{isele2019interactive} and (ii) a simpler \textit{probabilistic} method that classifies ``yielding'' behavior by projecting a relative acceleration onto a pre-determined Gaussian distribution of yielding \cite{wei2013autonomous}. Note, both models interact with one vehicle at a time (i.e., not interacting with multiple vehicles at once). That said, each interaction is time-efficient and a target vehicle of interaction is updated quickly (to address \textbf{[C3]} in Section~\ref{sec:introduction}).

We formally define that a lane-change is called ``Success'' if the ego vehicle moves to the target point (50 [m] ahead start point on the target lane) within a time-limit (80 $[s]$) without collisions\footnote{Note that, a collision can occur either by the ego vehicle crashing into a front vehicle or by a rear vehicle crashing into the ego vehicle.}. A throttle and angular acceleration are in the range of $[0,1]$, and a brake is in the range of $[-1,0]$, where $\pm 1$ indicates a maximum/minimum. Each model (i.e., NNMPC, Game Tree, and probabilistic) has a total of 50+ runs for each traffic scenario, and other vehicles are randomly positioned at each run.

TABLE~\ref{table:quantitative_analysis} reports the performance results for each traffic scenario. We highlight three findings from the analysis. First, the proposed method is robust to both the traffic density and cooperativeness of other drivers. Consequently, the $100\%$ success rate is consistently achieved in any traffic scenario. Second, the average brake, throttle, and angular acceleration indicate that a lane-change maneuver gets aggressive (yet not significant) with non-cooperative behaviors of other drivers and density of traffic. Third, compared to the baseline methods (game tree and probabilistic), NNMPC (Ours) also secures the smoother steering and takes 27\% shorter time\footnote{The absolute value of time may not represent the real-world driving behaviors, but the simulation environment represents extremely dense yet moving traffic to challenge lane-changing.} to merge. In contrast, the game tree model out-performs the smoothness in longitudinal planning (albeit not in lateral). In short, we emphasize that NNMPC secures the highest success rate with proactive coordination with other drivers (hence shorter time to merge), while securing smooth lateral maneuvers. 

It is also notable that NNMPC takes more time on average in the sparse cases (than the dense cases) and cooperative cases (than the aggressive cases). Recall that the ``Time'' in TABLE~\ref{table:quantitative_analysis} is the ``scenario completion time'' not ``time to merge'' (as defined above). Meaning, after the ego vehicle merges into the target lane, the ego vehicle has to slowly cruise with traffic until it passes the goal point (50 meters ahead of the starting point). Therefore, if the ego vehicle merges into the further inter-vehicle gaps at the front, the time will take less (it will be minimum if the ego vehicle merges ahead of the first vehicle in the queue). The results indicate that our method: (i) chooses one of the closer gaps from the starting point when other vehicles are cooperative and sparse (thus a longer completion time) and (ii) chooses one of the further gaps from the starting point when vehicles are aggressive and dense as the ego vehicle decides to cruise in the source lane and explore further gaps (thus a shorter completion time). We found that lane-changing generally takes more time when the ego vehicle stays and waits until enough space is found or other vehicles yield (while not). These behaviors are often found from the baseline methods (game tree and probabilistic), while our method proactively and successfully induces other vehicles to make room for the ego vehicle (consequently, less time while more aggressive brakes).

\subsubsection{Real-time Applicability of NNMPC}

The computation efficiency of the pipeline in Fig.~\ref{fig:diag-nnmpc} is dominated by the implementation of the neural networks. Thus, the choice of neural networks is crucial for the real-time applicability of this work. SGAN is known for its computational efficiency \cite{gupta2018social}, compared to its variants, e.g., S-LSTM \cite{alahi2016social}. The time complexity of SGAN corresponds to the matrix multiplications from LSTM and is asymptotically approximated to $\mathcal{O}(n^3)$ \cite{sundermeyer2012lstm} (although it can run even faster \cite{alman2021refined}). Given the input size $(N_p \times N_\text{veh})$, where $N_p = 8$ and $N_\text{veh} <= 3$ in our experiments, we evaluated the computation time in three different implementation settings: (i) Dell Optiplex 7070 Micro (6-Core, 9MB Cache, 2.2GHz to 3.7GHz, 35W, 8GB RAM), (ii) GT75 Titan 8RG (Intel CM246, GeForce GTX 1080, 64GB RAM), and (iii) Intel Xeon CPU ES-2640 v4 \@2.40GHz x 20, GeForce GTX TITAN. In summary, the computation time is approximated as 0.1 [sec] on average which was sufficient for the ROS communication with 10 Hz frequency. In particular, the first setting (i) is the on-board device for on-road driving and its computation time is 0.07 [sec] on average. The last setting (iii) is the setting we ran for the simulations, of which most cores were occupied by the Carla simulator and scenario runner. 
Recall that the computation efficiency highly depends on the choice of prediction modules as the computation time is dominantly consumed from iteratively evaluating a prediction module. Therefore, if a prediction module (not necessarily neural networks) is time-efficient, the overall computation time of the proposed pipeline can be as efficient. 
\begin{table*}[]
\centering
\caption{Quantitative/Comparative Analysis}
\begin{tabular}{@{}cccccccccccccccc@{}}
\toprule
\textbf{Model} & \textbf{Scenario} & \textbf{\begin{tabular}[c]{@{}c@{}}Num.\\ Runs\end{tabular}} & \textbf{Time} & \textbf{\begin{tabular}[c]{@{}c@{}}Succ.\\ (\%)\end{tabular}} & \textbf{\begin{tabular}[c]{@{}c@{}}Coll.\\ (\%)\end{tabular}} & \textbf{\begin{tabular}[c]{@{}c@{}}Time\\ out\\ (\%)\end{tabular}} & \textbf{\begin{tabular}[c]{@{}c@{}}Brake\\ Avg\end{tabular}} & \textbf{\begin{tabular}[c]{@{}c@{}}Thr.\\ Avg\end{tabular}} & \textbf{\begin{tabular}[c]{@{}c@{}}Acc.\\ Max\end{tabular}} & \textbf{\begin{tabular}[c]{@{}c@{}}Brake\\ Jerk\\ Avg\end{tabular}} & \textbf{\begin{tabular}[c]{@{}c@{}}Thr.\\ Jerk\\ Avg\end{tabular}} & \textbf{\begin{tabular}[c]{@{}c@{}}Ang.\\ Acc.\\ Avg\end{tabular}} & \textbf{\begin{tabular}[c]{@{}c@{}}Ang.\\ Acc.\\ Max\end{tabular}} & \textbf{\begin{tabular}[c]{@{}c@{}}Ang.\\ Jerk\\ Avg\end{tabular}} & \textbf{\begin{tabular}[c]{@{}c@{}}Ang.\\ Jerk\\ Max\end{tabular}} \\ \midrule
 & coop-sparse & 100 & \cellcolor[HTML]{34FF34}42.29 & \cellcolor[HTML]{34FF34}100 & \cellcolor[HTML]{34FF34}0 & \cellcolor[HTML]{34FF34}0 & -0.215 & 0.175 & 0.675 & -0.175 & 0.18 & \cellcolor[HTML]{34FF34}0.41 & \cellcolor[HTML]{34FF34}2.61 & \cellcolor[HTML]{34FF34}0.375 & \cellcolor[HTML]{34FF34}3.845 \\
 & coop-dense & 100 & \cellcolor[HTML]{34FF34}40.875 & \cellcolor[HTML]{34FF34}100 & \cellcolor[HTML]{34FF34}0 & \cellcolor[HTML]{34FF34}0 & -0.205 & 0.195 & 0.825 & -0.185 & 0.17 & \cellcolor[HTML]{34FF34}0.385 & \cellcolor[HTML]{34FF34}2.46 & \cellcolor[HTML]{34FF34}0.355 & \cellcolor[HTML]{34FF34}3.345 \\
 & agg-sparse & 100 & \cellcolor[HTML]{34FF34}41.065 & \cellcolor[HTML]{34FF34}100 & \cellcolor[HTML]{34FF34}0 & \cellcolor[HTML]{34FF34}0 & -0.18 & 0.175 & 0.765 & -0.165 & 0.15 & \cellcolor[HTML]{34FF34}0.41 & \cellcolor[HTML]{34FF34}2.705 & \cellcolor[HTML]{34FF34}0.37 & \cellcolor[HTML]{34FF34}3.145 \\
\multirow{-4}{*}{\textbf{Ours}} & agg-dense & 100 & \cellcolor[HTML]{34FF34}40.01 & \cellcolor[HTML]{34FF34}100 & \cellcolor[HTML]{34FF34}0 & \cellcolor[HTML]{34FF34}0 & -0.2 & 0.17 & 0.73 & -0.17 & 0.17 & \cellcolor[HTML]{34FF34}0.46 & \cellcolor[HTML]{34FF34}2.845 & \cellcolor[HTML]{34FF34}0.43 & \cellcolor[HTML]{34FF34}4.28 \\ \midrule
 & coop-sparse & 50 & 57.88 & 100 & 0 & 0 & \cellcolor[HTML]{34FF34}-0.1 & \cellcolor[HTML]{34FF34}0.11 & \cellcolor[HTML]{34FF34}0.58 & \cellcolor[HTML]{34FF34}-0.09 & \cellcolor[HTML]{34FF34}0.08 & 0.52 & 4.31 & 0.42 & 5.42 \\
 & coop-dense & 50 & 51.44 & 96 & 4 & 0 & \cellcolor[HTML]{34FF34}-0.11 & \cellcolor[HTML]{34FF34}0.08 & \cellcolor[HTML]{34FF34}0.53 & \cellcolor[HTML]{34FF34}-0.06 & \cellcolor[HTML]{34FF34}0.08 & 0.53 & 3.92 & 0.42 & 5.75 \\
 & agg-sparse & 50 & 48.87 & 96 & 4 & 0 & \cellcolor[HTML]{34FF34}-0.13 & \cellcolor[HTML]{34FF34}0.12 & \cellcolor[HTML]{34FF34}0.6 & \cellcolor[HTML]{34FF34}-0.1 & \cellcolor[HTML]{34FF34}0.09 & 0.57 & 4.32 & 0.46 & 5.46 \\
\multirow{-4}{*}{\begin{tabular}[c]{@{}c@{}}Game\\ Tree\end{tabular}} & agg-dense & 50 & 51.58 & 98 & 2 & 0 & \cellcolor[HTML]{34FF34}-0.14 & \cellcolor[HTML]{34FF34}0.07 & \cellcolor[HTML]{34FF34}0.53 & \cellcolor[HTML]{34FF34}-0.06 & \cellcolor[HTML]{34FF34}0.08 & 0.51 & 3.98 & 0.4 & 5.68 \\ \midrule
 & coop-sparse & 50 & 42.34 & 100 & 0 & 0 & -0.18 & 0.19 & 0.81 & -0.18 & 0.16 & 0.73 & 4.18 & 0.61 & 4.27 \\
 & coop-dense & 50 & 46.16 & 100 & 0 & 0 & -0.18 & 0.18 & 0.79 & -0.18 & 0.15 & 0.62 & 3.66 & 0.5 & 4.02 \\
 & agg-sparse & 50 & 41.21 & 96 & 4 & 0 & -0.18 & 0.2 & 0.91 & -0.18 & 0.16 & 0.86 & 4.65 & 0.74 & 5.44 \\
\multirow{-4}{*}{\begin{tabular}[c]{@{}c@{}}Probabi\\ -listic\end{tabular}} & agg-dense & 50 & 45.21 & 98 & 0 & 2 & -0.17 & 0.18 & 0.83 & -0.18 & 0.15 & 0.62 & 3.61 & 0.62 & 5.07 \\ \bottomrule
\end{tabular}
\label{table:quantitative_analysis}
\end{table*}

\subsubsection{Planning versus Control}

\begin{figure}
    \centering
    \includegraphics[width=1\columnwidth]{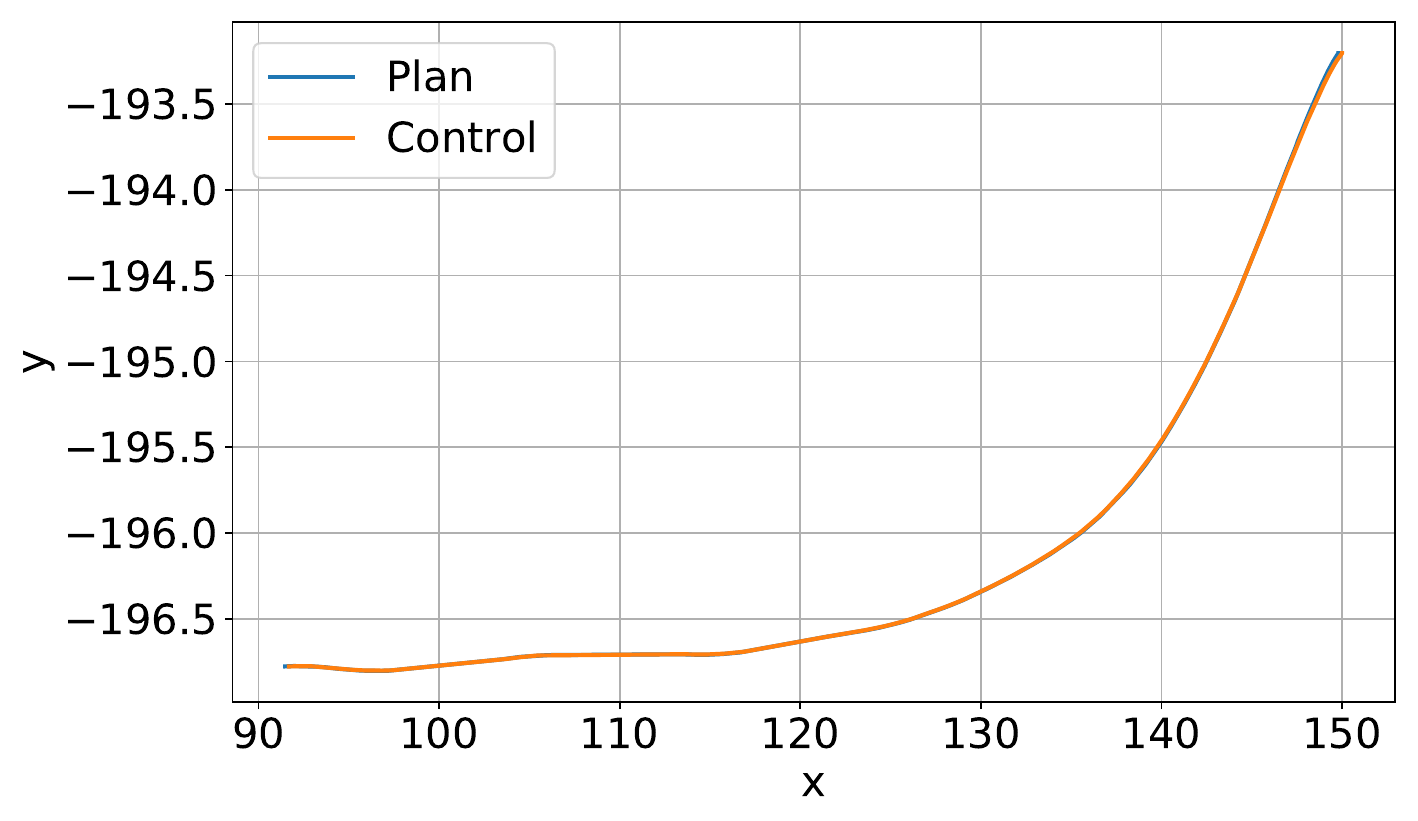}
    \vspace{-10pt}
    \caption{Planned versus executed trajectory. The planning trajectory only shows the first element of the entire trajectory, which the controller actuates at each time step.}
    \label{fig:plan_vs_control}
\end{figure}

Figure~\ref{fig:plan_vs_control} shows the planning trajectory from the planner and the actuated trajectory by the (speed and steering) controllers. It is clearly seen that the planning trajectory is well tracked by the controllers without delays. In fact, the controllers run much faster than the planner by a factor of two. The stability of the controllers is also mathematically proved in Appendix \ref{appendix:proof_of_stability}.

\subsubsection{Sensitivity Analysis on Noise}
To assess the impact of localization and perception noises on the performance of the proposed method, we investigate a sensitivity analysis with and without noises. The noise is set to 0.3 $[m]$ of standard deviation for both localization and perceptions. There are two main findings: (i) NNMPC is robust to perception noises while (ii) being susceptible to localization noises. 
NNMPC is robust to perception noises as SGAN was trained with noises in position. Consequently, the perception noises only marginally degrade the prediction accuracy. In fact, the localization noises also marginally degrade the prediction accuracy of SGAN for the same reason. However, the localization noises directly perturb the safety measure in \eqref{eq:const_collision}, causing either significant over-estimation or under-estimation. As a consequence, we see a more pronounced decrease of the success rate in TABLE \ref{table:noise_analysis}. The safety bound in \eqref{eq:safety_bound} can be jointly adaptive to localization noises, which remain for future work.

\begin{table}[]
\caption{Sensitivity analysis with localization and perception noise, in Aggressive-dense scenario.}
\label{table:noise_analysis}
\centering
\begin{tabular}{@{}ccc@{}}
\toprule
\multirow{2}{*}{\textbf{Noise Scale}} & \multicolumn{2}{c}{\textbf{Success Rate (\%)}}  \\
                                      & W/ Localization noise & W/   Perception noise \\ \midrule
1x & 95 & 100 \\
2x & 85 & 95  \\
5x & 60 & 95  \\ \bottomrule
\end{tabular}
\end{table}

\subsubsection{Performance at Edge Corner and Infeasible Solutions}
As mentioned, autonomous driving is one of the most safe-critical applications and thus the planner should be able to timely plan for corner cases, such as unexpected changes in the environment and stochastic behaviors of other traffic participants. In our lane changing scenario, the main edge case can be represented by incorrect evaluations on the other vehicles. The incorrectness is mainly with the failure of predictions, for example, by estimating a vehicle to slow down while it actually speeds up. Under dense traffic, the incorrect prediction may directly result in collisions. In our simulation runs, we occasionally observed the faulty predictions on the aggressive behaviors and consequently the ego vehicle had to return to the source lane (as illustrated in Fig.~\ref{fig:carla_result}). Another corner case could be the ego vehicle being stuck at the dead end (i.e., stopped vehicle) without having enough space to escape even with the max steering angle. Our method prevents the state of being stuck successfully with the feasible set in \eqref{eq:feas_set}, as opposed to our baseline (probabilistic) that occasionally gets stuck under dense and aggressive traffic conditions.

Although our method shows robustness to these corner cases, the scenario is relatively simple and accordingly corner cases are limited. Thus, validating algorithms under more complex scenarios with different types/behaviors of traffic participants remain for future work.

Also, it is important to point out that Eqn.~\eqref{eq:opt} can be infeasible. In particular, the safety constraint in Eqn.~\eqref{eq:const_collision} can be violated depending on the safety bound size $\epsilon$. To avoid the infeasibility, we can reformulate the hard constraint Eqn.~\eqref{eq:const_collision} as a soft constraint and add it as an additional penalty in the objective function \eqref{eq:obj}, i.e.,
\begin{equation}
    +\lambda_{g}\sum_{\ell=t}^{t+T}\|g_i (z(\ell+1|t);x_i(t),y_i(t))-\epsilon\|^2.
\end{equation}
Now, the safety constraint is not hard and therefore the feasibility associated with it can be guaranteed. With this soft constraint, the problem \eqref{eq:opt} is feasible with all candidates and will choose the best candidate. Depending on the optimality of the trajectory candidates, the best solution could be either safe or colliding with other vehicles. In this specific case of a scenario where the traffic is very slow ($\sim$3m/s), we consider that ``stopping and waiting" is a good strategy that reflects real-world driving behaviors. Therefore, we intentionally remained the safety constraint as hard, so that when it does not seem to be safe, the ego vehicle stops and waits. Nevertheless, there could be alternatives when solutions become infeasible and the examples include: (i) increasing the speed reference or (ii) decreasing the speed reference, which enables explorations of other gaps. It is also possible to consider these exploring trajectories as additional ``trajectory candidates" with extra computational expenses, which is left to a control engineer as a design choice.

\subsection{Limitation, Extension, and Future Work}
The primary limitation of this study is related to its validation. Recall that the validation is under the recognized CARLA simulation with the practical implementation setup for planning and control (using ROS nodes). However, traffic vehicles are simulated using IDM, where the driving maneuvers may significantly differ from that of real vehicles on road. Furthermore, our validation scenarios are with the extremely dense traffic (for both ``sparse'' and ``dense'' cases) that may not represent real-world scenarios (where a ``free space'' occasionally exists).
As a next step, we plan to incorporate a human driver in the simulation to drive one of the neighboring vehicles and to validate it under diverse scenarios (including moderate traffic scenarios) to reduce sim-to-real gaps. We are also in the progress of running on-road experiments at a test facility in Japan.\\
\indent The second limitation is related to safety. Although safety is not guaranteed in dense traffic, there exist several approaches to enhancing safety: (i) improving the prediction accuracy of other vehicles' cooperative behavior (as shown in \cite{bae2019cooperation}), (ii) integrating distributionally robust methods \cite{kandel2020distributionally} that learns prediction errors online, (iii) similarly, integrating a distribution of predicted positions instead of single position prediction, and (iv) dynamic adjustments of a selected candidate based on predictions/errors, e.g., reducing a steering angle to increase the inter-vehicle gap to a rear vehicle. These are under active research.\\
\indent Lastly, the optimal solutions from Eqn.~\eqref{eq:smooth_path} do not necessarily represent the optimal solutions from Eqn.~\eqref{eq:opt}, i.e., the optimal solution in Eqn.~\eqref{eq:smooth_path} may be suboptimal in Eqn.~\eqref{eq:opt}. Recall, the optimization in Eqn. \eqref{eq:opt} is mainly to evaluate the safety (with predictions) while the optimization in Eqn.~\eqref{eq:smooth_path} is to generate smooth maneuvers for each intent. That said, the optimality of Eqn.~\eqref{eq:opt} can enhance by generating multiple candidates for each intention. For example, for the lane-change intent, three maneuvers can be generated to represent immediate-change, move-forward-and-change, and slow-down-and-change. Generating multiple maneuvers for each intent also remains for future work. All that being said, the proposed method is a framework that combines optimization-based models with data-driven models (e.g., neural networks) which is not limited to a specific choice of methods.

\section{Conclusions}\label{sec:conclusion}
This paper presents mathematical formulations for smooth lane-change control in dense traffic where inter-vehicle gaps are narrow. We build upon an MPC formulation for trajectory planning with moving objects, by integrating a state-of-the-art Recurrent Neural Network, Social Generative Adversarial Networks (SGAN) to estimate interactions between objects -- together being called NNMPC. NNMPC generates smooth lane-change trajectories to improve passenger comfort. In particular, possible driving intentions are specified and refined to control candidates. Additionally, sensor noises, SGAN prediction errors, and recursive feasibility for control are incorporated to increase robustness. The simulation studies with CARLA validate the smoothness of the trajectory and computational efficiency as well as the robustness to traffic density and cooperativeness of other drivers.

\section*{Acknowledgement}
This paper is an extension of work \cite{bae2019cooperation} presented at American Control Conference, Denver, USA, in 2020. The authors thank Donggun Lee at the University of California, Berkeley, for valuable discussions.

\appendices

\bibliographystyle{IEEEtran}
\bibliography{refs}


%

\appendix[Proof of Control Stability]\label{appendix:proof_of_stability}
We refer to \cite{wen1990pid} for the stability of the PID controller (for the speed control) and we hereby show the stability of the steering control which is based on MPC. MPC with constraints (i.e., the nonlinear vehicle rotation model and curvature constraints from \cite{tashiro2013vehicle}) does not typically guarantee stability and requires significant efforts to find a Lyapunov function \cite{valluri1998stability} -- without any guarantee of existence. Thus, we show that the deviation of the steering control with respect to a reference is bounded. Also, recall that the reference trajectory is obtained from \eqref{eq:smooth_path}, not from \eqref{eq:opt}.
\begin{proposition}

Given a feasible solution from \eqref{eq:smooth_path}, the maximum deviation of steering control $\tilde{\delta}$ from reference $\delta^\text{ref}$ is bounded by a finite value $K$ over a control horizon $T$, i.e.,
\begin{equation}
    \max_{\tau \in \{0,\ldots,T\}}\;\|\delta^\text{ref}(\tau)-\tilde{\delta}(\tau)\|\leq K
\end{equation}
\end{proposition}
\textit{Proof.} At any time step $\tau \in \{0,\ldots,T\}$, the reference steering angle from \eqref{eq:smooth_path} is bounded by $\mathcal{K}_\text{max}$, i.e., $-\mathcal{K}_\text{max}\leq\delta(\tau)\leq\mathcal{K}_\text{max}$ (The reference trajectory is feasible and therefore \eqref{eq:curv_p1} and \eqref{eq:curv_p2} must suffice).
Now, at any time step $\tau$, the maximum deviation is achieved when the curvature constraints \eqref{eq:curv_p1} and \eqref{eq:curv_p2} are active, i.e., $\delta^\rf = \mathcal{K}_\text{max}$ or $\delta^\rf = -\mathcal{K}_\text{max}$. And we will check if the deviation at each case is bounded.
First, if $\delta^\rf = \mathcal{K}_\text{max}$,
\begin{align}
    \max_{\tau \in \{0,\ldots,T\}}\;\|&\delta^\rf(\tau)-\tilde{\delta}(\tau)\|\\
        &=\max_{\tau \in \{0,\ldots,T\}}\;\|\mathcal{K}_\text{max}-\tilde{\delta}(\tau)\|\\
        &\leq\mathcal{K}_\text{max} + \max_{\tau \in \{0,\ldots,T\}} \|-\tilde{\delta}\| \quad(\because \mathcal{K}_\text{max} > 0)\\
        &=\mathcal{K}_\text{max} + \tilde{\delta}_\text{max}.
\end{align}
Similarly, if $\delta^\rf = -\mathcal{K}_\text{max}$, 
\begin{align}
    \max_{\tau \in \{0,\ldots,T\}}\;\|&\delta^\rf(\tau)-\tilde{\delta}(\tau)\|\\
        &=\max_{\tau \in \{0,\ldots,T\}}\;\|-\mathcal{K}_\text{max}-\tilde{\delta}(\tau)\|\\
        &\leq\mathcal{K}_\text{max} + \max_{\tau \in \{0,\ldots,T\}} \|-\tilde{\delta}\| \quad(\because \mathcal{K}_\text{max} > 0)\\
        &=\mathcal{K}_\text{max} + \tilde{\delta}_\text{max}.
\end{align}
Therefore, the following suffices
\begin{equation}
\max_{\tau \in \{0,\ldots,T\}}\;\|\delta^\text{ref}(\tau)-\tilde{\delta}(\tau)\|\leq \mathcal{K}_\text{max} + \tilde{\delta}_\text{max} = K,
\end{equation}
which concludes the proof. $\blacksquare$

\begin{IEEEbiography}[{\includegraphics[width=1in,height=1.25in,clip,keepaspectratio]{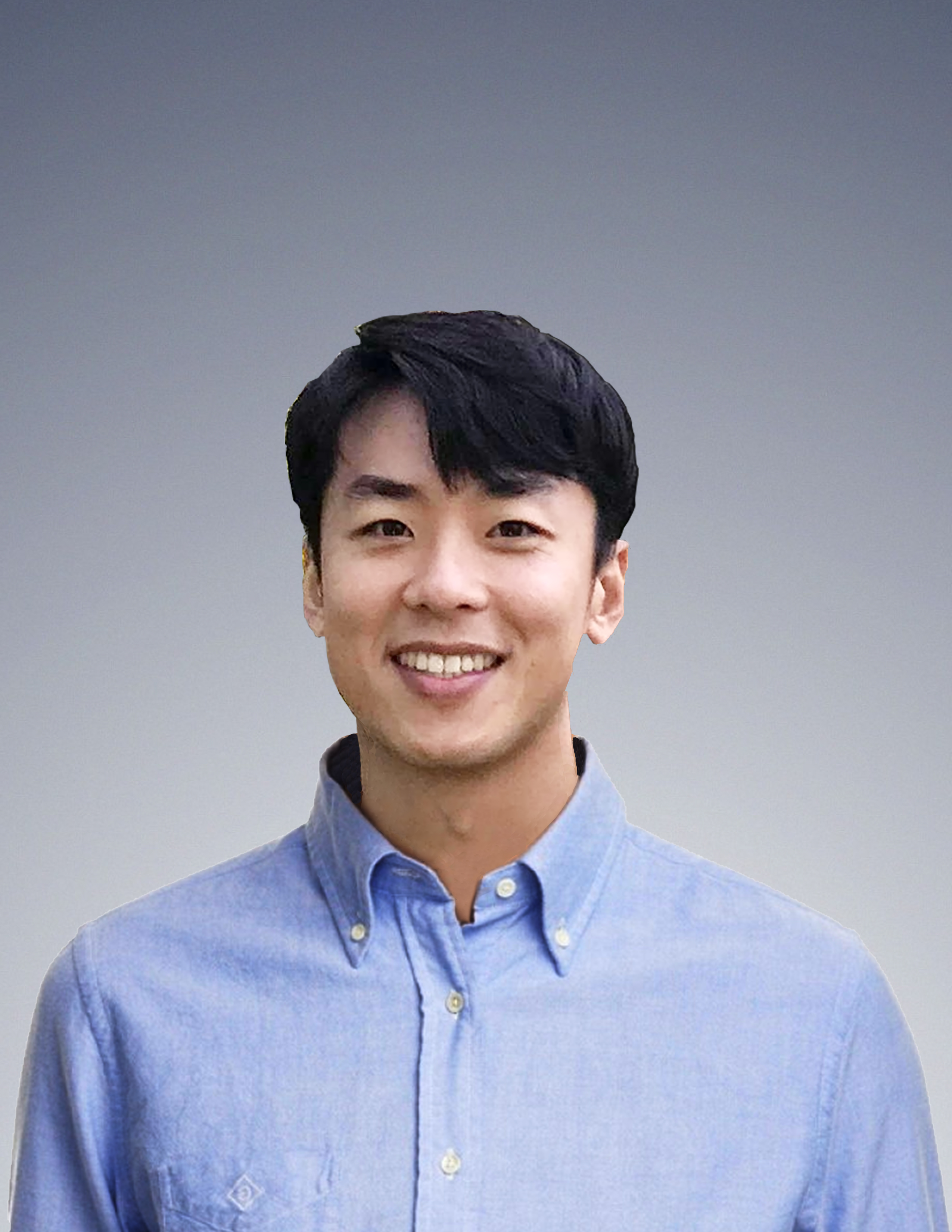}}]
{Sangjae Bae}
received the B.S. degree in electronics engineering from Kyung Hee University, Korea, in 2015, and received the Ph.D. degree in systems engineering from the University of California, Berkeley, in 2020.

Dr. Bae is currently a senior scientist at Honda Research Institute, USA, where he works on applications of optimization, control, and machine learning, with a focus on autonomous vehicles. 

He was nominated for the IFAC Young Author Award at the 2nd IFAC Workshop on Cyber-Physical and Human Systems in 2018.
\end{IEEEbiography}

\begin{IEEEbiography}[{\includegraphics[width=1in,height=1.25in,clip,keepaspectratio]{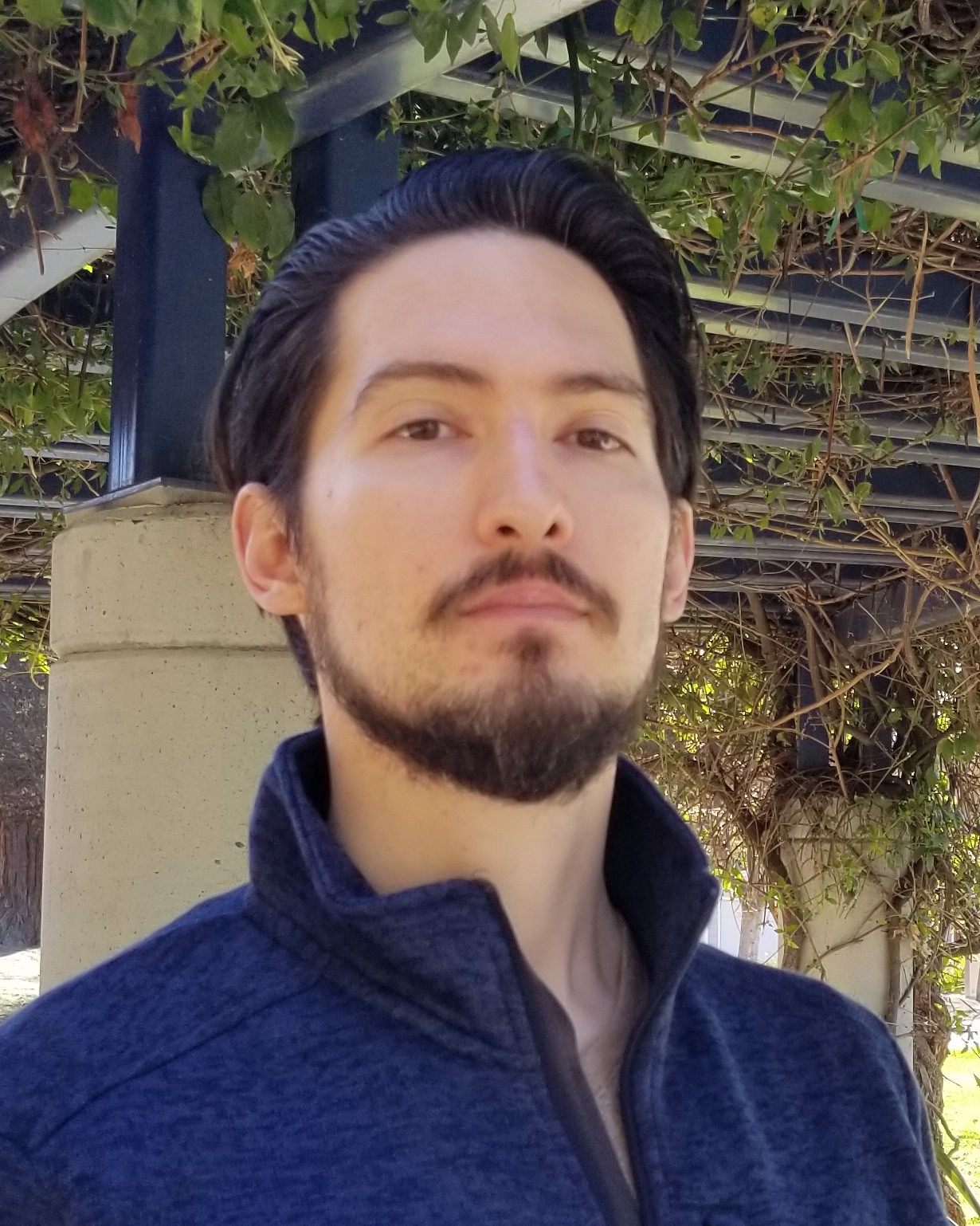}}]{David Isele}
received his B.E. degree in electrical engineering from The Cooper Union: Albert Nerken School of Engineering, in New York City. He received his M.S.E. degree in Robotics and his Ph.D. degree in Computer and Information Science from The University of Pennsylvania.  
	
David is currently a senior scientist at Honda Research Institute US. His research interests include applications of machine learning and artificial intelligence to robotic systems, with a focus on strategic decision making for autonomous vehicles. 

\end{IEEEbiography}


\begin{IEEEbiography}[{\includegraphics[width=1in,height=1.25in,clip,keepaspectratio]{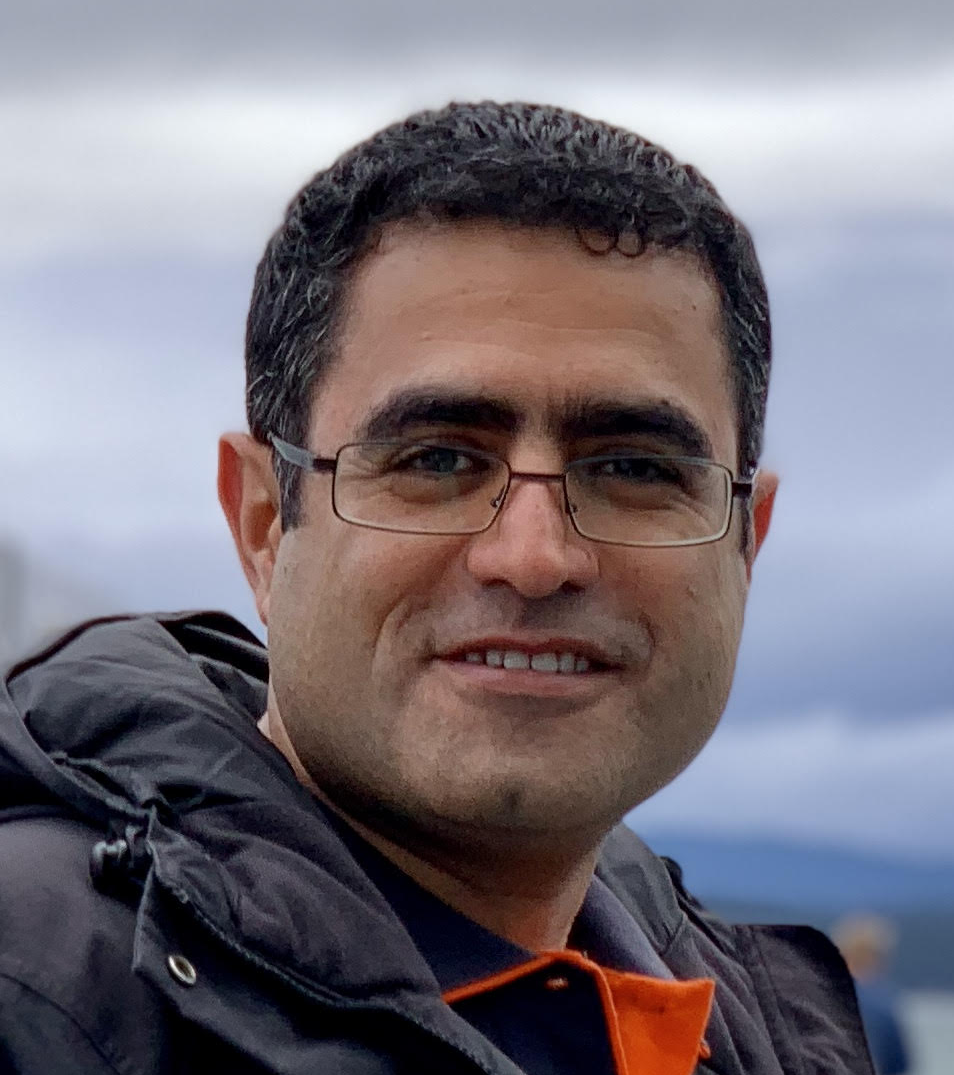}}]{Alireza Nakhaei}
received his PhD degree in computer science and robotics from LAAS-CNRS in Toulouse, France.

Alireza's research interests include employing artificial intelligence in decision making algorithms for self-driving vehicle and leveraging the autonomous driving capabilities to improve the performance of driver-assistance systems (ADAS).  

\end{IEEEbiography}
\begin{IEEEbiography}[{\includegraphics[width=1in,height=1.25in,clip,keepaspectratio]{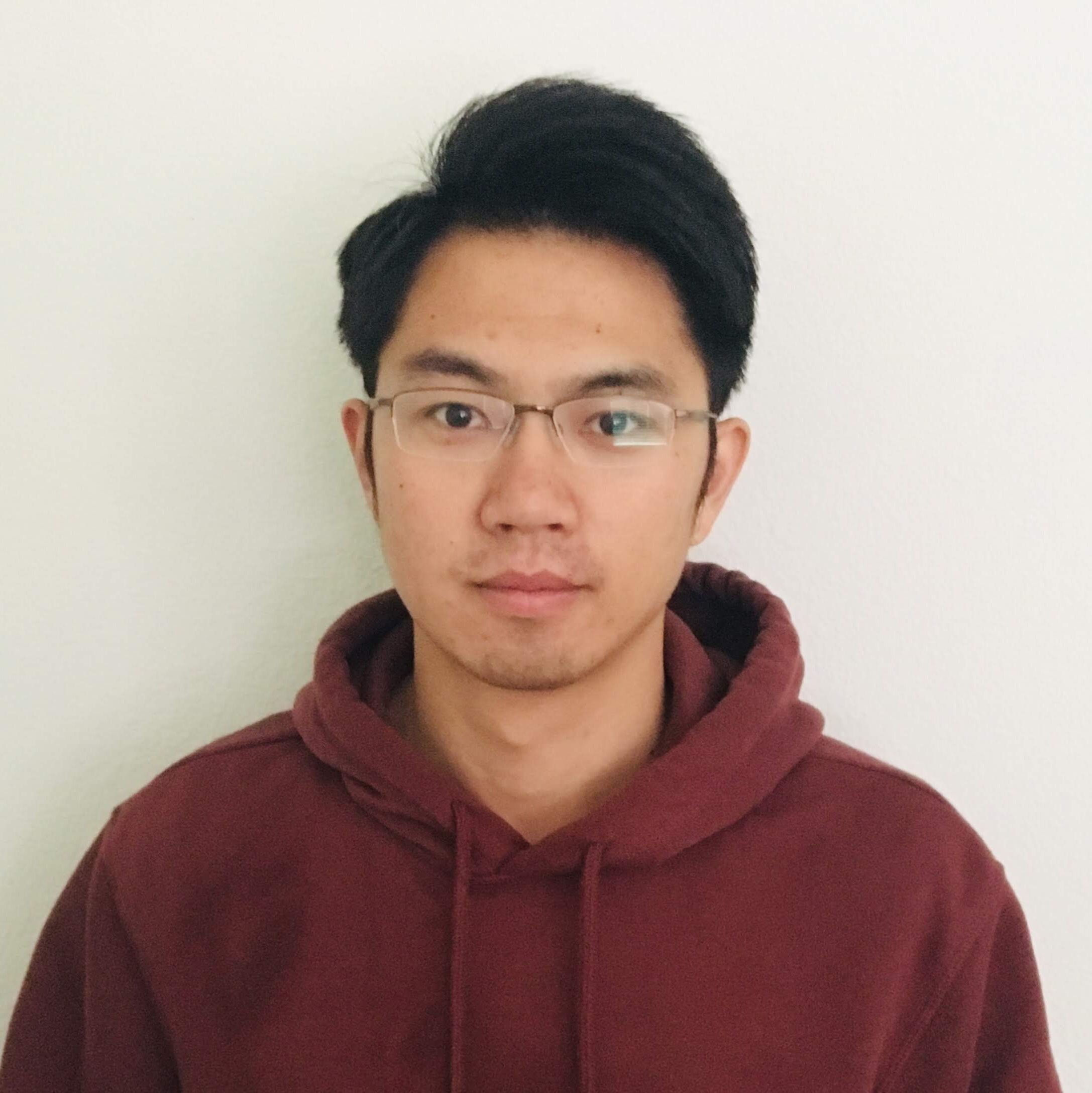}}]{Peng Xu}
 received his B.E. degree in Aircraft Power Engineering from Nanjing University of Aeronautics and Astronautics, Nanjing, China, in 2010 and his Master degree in Naval Architecture and Ocean Engineering from Shanghai Jiao Tong University, Shanghai, China in 2013. He also received his second Master degree in Electrical Engineering from Case Western Reserve University, Ohio in 2018.

His research interests include applying probabilistic methods, optimization based methods and learning based methods on decision making and motion planning for autonomous vehicles and mobile robots.

\end{IEEEbiography}

\begin{IEEEbiography}[{\includegraphics[width=1in,height=1.25in,clip,keepaspectratio]{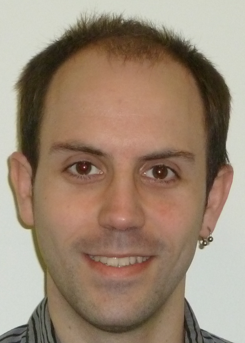}}]{Alexandre Miranda A$\tilde{\text{n}}$on}
received his Dipl. Ing degree from the School of Industrial Engineering of Barcelona (ETSEIB-UPC), Catalonia. He received his M.S.E. degree in Robotics from the University of Pennsylvania.

Alex is currently a Research Engineer and the Assistant Lead Engineer for the System Engineering team at Honda Research Institute USA. His interest include the use of artificial intelligence for robotics, autonomous systems and self-driving vehicles with a focus on decision making and motion planning algorithms.
\end{IEEEbiography}

\begin{IEEEbiography}[{\includegraphics[width=1in,height=1.25in,clip,keepaspectratio]{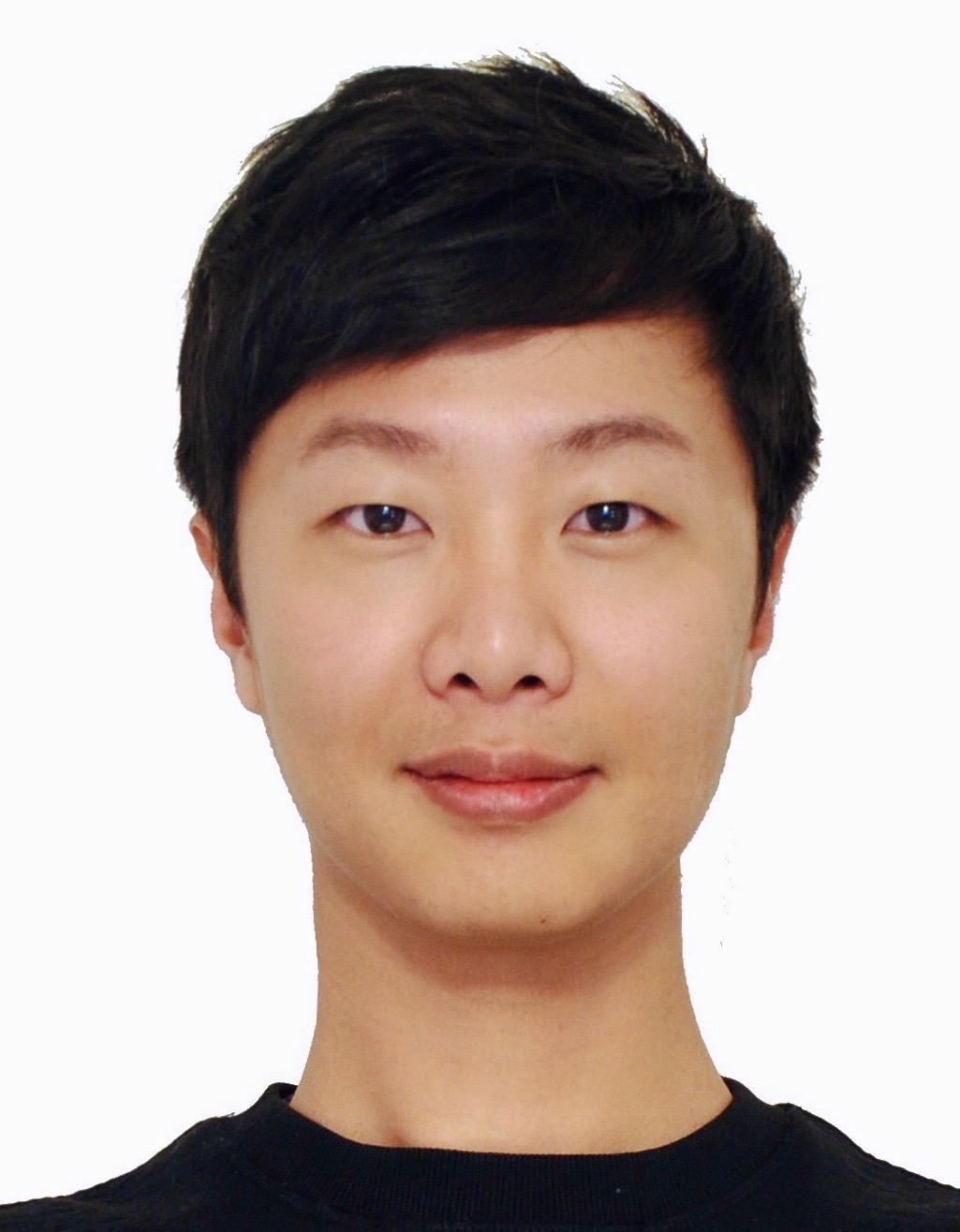}}]{Chiho Choi}
received a B.S. from Hanyang University, Korea, and a M.S. in electrical engineering from University of Southern California. He received his Ph.D. in electrical and computer engineering from Purdue University in 2018.

Dr. Choi is currently a senior scientist at Honda Research Institute USA. His research interests span the fields of computer vision, machine learning, and robotics with a focus on understanding and prediction of human behavior for the safe operation of vehicles and robots designed to cooperate with humans.
\end{IEEEbiography}

\begin{IEEEbiography}[{\includegraphics[width=1in,height=1.25in,clip,keepaspectratio]{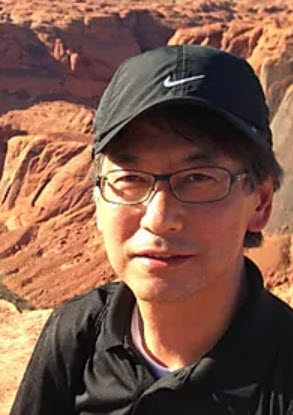}}]{Kikuo Fujimura}
received the B.S. and M.S. degrees in Information Science from the University of Tokyo and the Ph.D. degree in Computer Science from the University of Maryland, College Park in 1989.  After working at Oak Ridge National Laboratory and Ohio State University (Columbus), he joined Honda R\&D in 1998, where he was engaged in research on intelligent systems and human-robot interaction with Honda’s humanoid robot ASIMO.   

He is currently Director Innovation at Honda Research Institute USA in San Jose, California, where he directs teams of researchers working on automated driving, knowledge discovery and informatics, human machine interfaces, and intelligent robotics. His research interests include artificial intelligence for mobility, human robot interaction, and HCI. 

He has authored/co-authored one book and over 100 research papers in refereed conferences and journals and has been granted over 20 patents. He currently serves as an Associate Editor of IEEE Transactions on Intelligent Vehicles.

\end{IEEEbiography}

\begin{IEEEbiography}[{\includegraphics[width=1in,height=1.25in,clip,keepaspectratio]{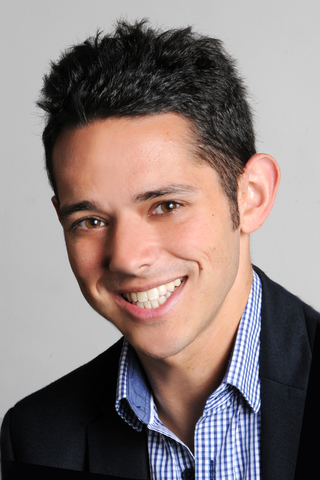}}]{Scott Moura} 
received a B.S. degree (2006) from the University of California, Berkeley, and M.S. (2008) and Ph.D. (2011) degrees from the University of Michigan, Ann Arbor. He was a Postdoctoral Fellow at the University of California, San Diego in 2011-2013. 

Dr. Scott Moura is currently the Clare and Hsieh Wen Shen Endowed Distinguished Professor, Director of eCAL, PATH Faculty Director, and Chair of Engineering Science at UC Berkeley. From 2015 – 2021 he was PI at the Tsinghua-Berkeley Shenzhen Institute. In 2013, he was a Visiting Researcher at MINES ParisTech, France. His research includes control, optimization, and data science for batteries, electrified vehicles, and climate-tech. 

Dr. Moura has received the ASME Dynamic Systems and Control Division Young Investigator Award, National Science Foundation CAREER Award, Carol D. Soc Distinguished Graduate Student Mentor Award, O. Hugo Shuck Best Paper Award,  UC Presidential Postdoctoral Fellowship, and NSF Graduate Research Fellowship.



\end{IEEEbiography}




\end{document}